\documentclass[prx,aps,showpacs,lengthcheck,superscriptaddress,twocolumn]{revtex4}
\usepackage{amsmath}
\pdfoutput=1
\usepackage{xcolor}
\usepackage{amsfonts}
\usepackage{graphicx}
\usepackage{bm}

\newcommand{\beq}{\begin{equation}}
\newcommand{\eeq}{\end{equation}}
\newcommand{\beqa}{\begin{eqnarray}}
\newcommand{\eeqa}{\end{eqnarray}}

\begin{document}

\title{Correlating Cell Shape and Cellular Stress in Motile Confluent Tissues}

\author{Xingbo Yang}
\affiliation{Department of Physics, Northwestern University, Evanston, Illinois 60208, USA}
\author{Dapeng Bi}
\affiliation{Department of Physics, Northeastern University, Boston, MA 02115, USA}
\author{Michael Czajkowski}
\affiliation{Physics Department and Syracuse Soft Matter Program, Syracuse University, Syracuse NY 13244, USA}
\author{Matthias Merkel}
\affiliation{Physics Department and Syracuse Soft Matter Program, Syracuse University, Syracuse NY 13244, USA}
\author{M. Lisa Manning}
\affiliation{Physics Department and Syracuse Soft Matter Program, Syracuse University, Syracuse NY 13244, USA}
\author{M. Cristina Marchetti}
\affiliation{Physics Department and Syracuse Soft Matter Program, Syracuse University, Syracuse NY 13244, USA}

\begin{abstract}
{Collective cell migration is a highly regulated process involved in wound healing, cancer metastasis and morphogenesis. Mechanical interactions among cells provide an important regulatory mechanism to coordinate such collective motion. Using a Self-Propelled Voronoi  (SPV) model that links cell mechanics to cell shape and cell motility, we formulate a generalized mechanical inference method to obtain the spatio-temporal distribution of cellular stresses from measured traction forces in \textit{motile} tissues and show that such traction-based stresses match those calculated from instantaneous cell shapes. 
We additionally use stress information to characterize the rheological properties of the tissue. We identify a motility-induced swim stress that adds to the interaction stress to determine the global contractility or extensibility of epithelia. We further show that  the temporal correlation of the interaction shear stress determines an effective viscosity of the tissue that diverges at the liquid-solid transition,  suggesting the possibility of extracting  rheological information directly from traction data.
}

\end{abstract}

\maketitle

It is now broadly recognized that the transmission of mechanical forces can be as important as genetics and biochemistry in regulating tissue organization in many developmental processes, including embryogenesis, morphogenesis, wound healing and cancer metastasis~\cite{Lecuit2007,Bellaiche2013,Gardel2015,Eaton2015,Shraiman2005,Xavier2012,Staple2010,Guillot2013,Lenne2008,Xavier2009,Dhananjay2011}. In order to make quantitative predictions for large scale cell remodeling in tissues, we must understand their material properties, such as stiffness and viscosity, as well as the forces that build up inside them, characterized by local pressures and stresses.
Motivated by experiments highlighting the slow glassy dynamics of dense epithelial tissues, work by us and others has suggested that monolayers of motile cells may form glassy or jammed states, where local cell rearrangements are rare and energetically costly, and that
a relatively small change of parameters may trigger a change from an elastic response to a state with fluid-like
behavior, where individual cells continuously exchange neighbors~\cite{Bi2014,Manning2015,Bi2016}. This liquid-solid transition is tuned by the interplay of cell-cell adhesion and cortex contractility, as manifest in cellular shape, and by cell motility. This suggestion has been verified experimentally in specific cell types~\cite{Park2015}, indicating that the paradigm of tissues as active materials may be a useful way of organizing experimental data  and classifying large-scale tissue behavior in terms of a few effective parameters. 

In contrast, quantifying stresses and pressures in active, motile tissues is largely an open problem.  One possible reason  is that there are different definitions of stresses and pressures that arise naturally in different experiments and simulations, and it is not immediately clear how they are related to one another or under precisely which conditions each definition applies.

For example, Traction Force Microscopy (TFM) is a powerful tool that probes the dynamic forces cells exert onto soft substrates by measuring the substrate deformations~\cite{Dhananjay2011,Gardel2014,Dhananjay2013,Zimmerman2014,Xavier2009}. In some experiments, intercellular stresses are extracted from the traction forces using finite element analysis under the condition that traction forces are balanced by cellular interactions. The resulting stress maps reveal a highly dynamical and heterogeneous mechanical landscape, characterized by large spatial and temporal fluctuations in both normal and shear stresses~\cite{Dhananjay2011,Dhananjay2013,Jae2013,Xavier2009,Xavier2012}. A key assumption used in the TFM approach to infer stresses from tractions is that the cell layer can be described as a continuum linear elastic material~\cite{Dhananjay2013}. 

A second important set of \textit{mechanical inference} methods also predict cellular stresses, but in static tissues. These methods rely on advances in imaging techniques that provide a spatially resolved view of tissue development during morphogenesis, with visualization of the cell boundaries of two dimensional cell sheets~\cite{Ishihara2012,Marcp2016,Kevin2012,Noll2015,Brodland2010}. Assuming mechanical equilibrium, one can then infer the tensions along cell edges and pressures within each cell from the cell configurations. This method provides a spatial distribution of intercellular stresses, but it does not capture temporal stress fluctuations arising from the nonequilibrium nature of the tissue. On the other hand, it directly couples  mechanics to morphology and it has been used successfully  to characterize cell morphology in the development of the {\it{Drosophila}} wing disk~\cite{Farhadifar2007,Kevin2012}, where cellular rearrangements  are slow on the time scales of interest. This work has demonstrated that the analysis of cell shapes can provide fundamental insight on the mechanical state of tissues in developmental processes. 

 A third line of research has focused on the {\em homeostatic pressure} that tissues exert on their containers or surroundings. The homeostatic pressure has been proposed as a quantitative measure of the metastatic potential of a tumor~\cite{Prost2009,Montel2011}.  It is defined as the force per unit area that a confined tissue would exert on a moving piston permeable to fluid, hence it represents an active osmotic cellular pressure. The existing literature has focused on the contribution to the homeostatic pressure from tissue growth due to cell division and death, but in general other active processes, such as cell motility and contractility, will also contribute to the forces exerted by living tissues on confining walls.  A related body of work has investigated the pressures generated by motile particles, such as active colloids. In these highly nonequilibrium systems that break time-reversal symmetry, there is a contribution to the total pressure called the {\em swim pressure} generated entirely by persistent motility~\cite{XY2014,Takatori2014,Solon2015}.

\begin{figure}[!h]
\begin{centering}
\includegraphics[width=1.00\columnwidth]{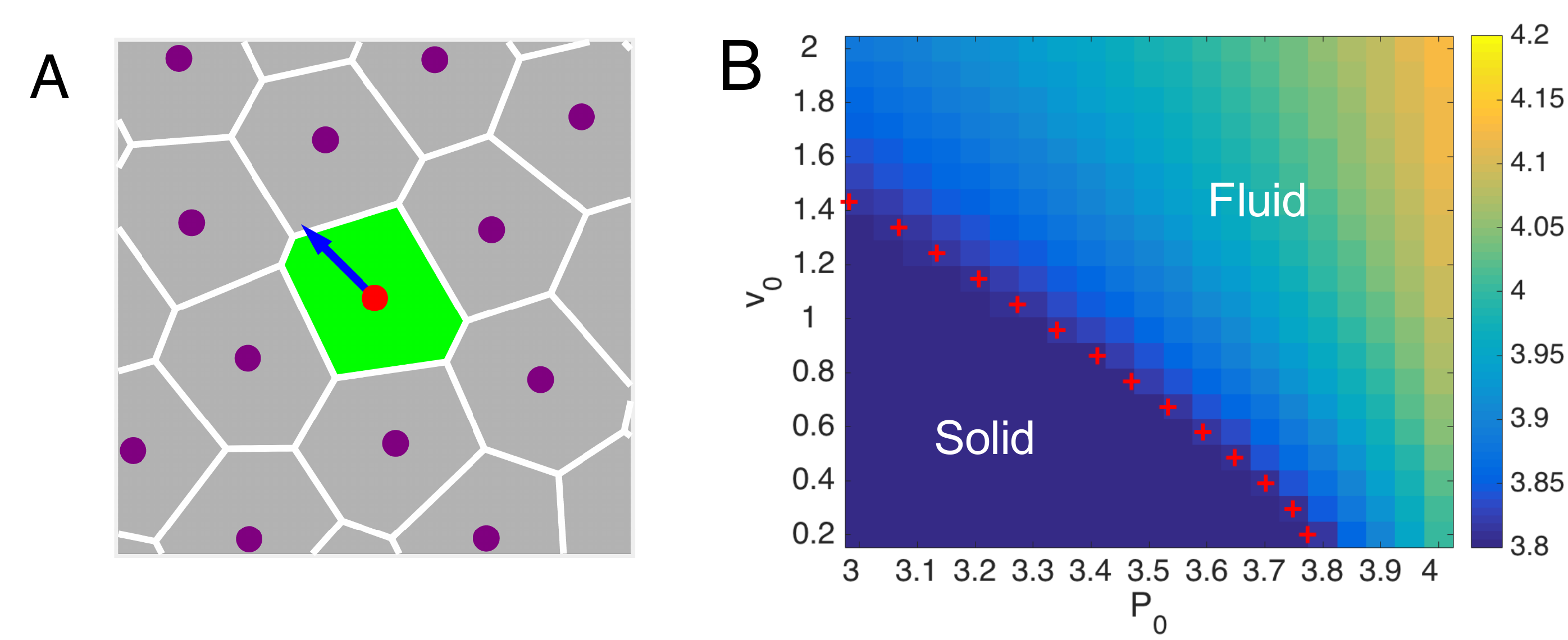}
\caption{A: Illustration of the SPV model, where cells are represented by polygons obtained via a Voronoi tessellation of initially random cell positions, with a self-propulsion force applied at each cell position. B: Phase diagram in the $(P_0,v_0)$ plane based on the value of the shape parameter $q$ (color scale), with the phase boundary (red crosses) determined by $q=3.813$.}
\label{fig:spv2}
\end{centering}
\end{figure}

So far, there is no unifying theory for these seemingly distinct notions of pressure and stress, or their relationship to material properties. In this paper, we show that a recently proposed Self-Propelled Voronoi (SPV) model of epithelial tissue (Fig.~\ref{fig:spv2}) provides a natural framework for unifying these ideas.  In the SPV model, cells are described via a polygonal tesselation of the plane. Each cell is endowed with motility with fixed speed and direction randomized by orientational noise and interacts with other cells via forces determined by a shape energy that incorporates the interplay between contractility and adhesion. 
In a broad class of vertex models~\cite{Farhadifar2007} one can develop a self-consistent definition for cellular stress in terms of cell shapes, but this result is only useful in simulations where the energy functional that describes the tissue is precisely known.  One of the benefits of the SPV model is that it additionally explicitly accounts for the forces that motile cells exert on the substrate.
This allows us to develop a generalized mechanical inference method to infer cellular stresses from traction forces and to show that these match the stresses calculated from instantaneous cell shape, relating for the first time TFM data and mechanical inference techniques in motile tissues. Additionally,  our method provides absolute values for junctional tensions and pressure differences. This is in contrast to equilibrium mechanical inference, which only yields relative forces~\cite{Kevin2012,Ishihara2012}.

We also wish to know how this definition of cellular stress relates to others in the literature. There are two additive contributions to the mechanical stress that describe the forces transmitted in a material across a bulk plane. The first represents the flux of propulsive forces through a bulk plane carried by particles that move across it. The second describes the flux of interaction forces across a bulk plane.  We demonstrate that the generalized mechanical inference measurements probe the latter, which we denote as {\em interaction stresses}. The former, which we denote as the tissue {\em swim stress}, approximates the contribution from cell motility to the osmotic pressure generated by cells immersed in a momentum conserving solvent on a semipermeable piston,
hence to the tissue homeostatic pressure. The tensorial sum of the swim stress and the interaction stress is the {\em total stress}. The normal component of the total stress determines whether a tissue will tend to exert extensile or contractile forces on its environment, which is an important consideration in biological processes such as wound healing and cancer tumorogenesis.

An obvious open question, then, is how these stresses vary as a function of material properties. We find that the normal component of the interaction stress is contractile in both the solid and the liquid due to the contractility of the actomyosin cortex, although much more weakly so in the liquid state.  In contrast, the normal component of the motility-induced swim stress is always extensile, corresponding to a positive swim pressure, though its magnitude depends on the phase: in a solid the swim pressure is negligible, while in the fluid it can be significant. This can result in a change in sign of the total mean stress:
indeed, we find it is always contractile in the solid state but becomes extensile deep in the liquid state when cell motility exceeds actomyosin contractility.

Because the transition from contractile to extensile does not coincide with the fluid to solid transition, it is natural to ask whether the stress displays any signatures of the fluid-solid transition.  We develop a definition for the effective viscosity of the tissue that can be extracted from the temporal correlation of the interaction shear stress, and find that it diverges as the tissue transits from the liquid state to the solid state. 
Importantly, this suggests that
TFM combined with mechanical inference can provide rheological information about the tissue. 
\section*{Results and Discussion}
\paragraph*{Self-Propelled Voronoi (SPV) Model.} 
The SPV model describes an epithelium as a network of polygons.  Each cell $i$ is endowed a position vector $\bm{r}_i$, and cell shape is defined by the Voronoi tessellation of all cell positions  (Fig.\ref{fig:spv2}). Like for vertex models, tissue forces are obtained from an effective energy functional $E(\lbrace\bm{r}_i\rbrace)$ for $N$ cells, 
given by~\cite{Farhadifar2007,Staple2010,Bi2016,Rastko2016}
\begin{gather}
E=\sum_{i=1}^N E_i\;,~E_i=K_A(A_i-A_0)^2+K_P(P_i-P_0)^2\;,
\label{eqn:energy_square}
\end{gather}
with $A_i$ and $P_i$ the cross-sectional area and perimeter of the $i$-th cell, respectively. The first term in Eq.~\eqref{eqn:energy_square} arises from incompressibility of the layer in three dimensions and its resistance to height fluctuations, with $A_0$ a preferred cross-sectional area. The second term represents the competition between cortical tensions from the actomyosin network at the apical surface and cell-cell adhesions from adhesive complexes at intercellular junctions~\cite{Farhadifar2007}, with $P_0$ a preferred perimeter resulting from this competition.
We simulate $N$ cells in a square box of area $A_{T}$, with $\bar{A}=A_T/N$ the average cell area and with periodic boundary conditions.  The system is initialized with a set of $N$ random cell positions, independently drawn from a uniform distribution.
Throughout the simulations, we set $\bar{A}=A_0=1$ unless otherwise noted, and $K_A=K_P=1$.

Each Voronoi cell is additionally endowed with a constant self-propulsion speed $v_0$ along the direction of polarization $\hat{\bm n}_i=(\cos\theta_i,\sin\theta_i)$ describing cell motility.  The dynamics of each Voronoi cell is governed by 
\begin{equation}
\partial_t \bm r_i = \mu \bm F_i +  v_0 \hat{\bm n}_i\;,\hspace{0.2in}\partial_t \theta_i = \sqrt{2D_r}\eta_i(t)\;,
\label{equation_of_motion_SPV}
\end{equation}
where $\bm F_i=-\bm \nabla_i E$ is the force on cell $i$ and $\mu$ is the mobility. The  direction of cell polarization is randomized by orientational noise of rate $D_r$, with $\langle\eta_i(t)\rangle=0$ and $<\eta_i(t)\eta_j(t')>=\delta_{ij}\delta(t-t')$. The time scale $\tau_r=1/D_r$ controls the persistence of single-cell dynamics. As in Self-Propelled Particle (SPP) models, an isolated cell performs a persistent random walk with a long-time translational diffusivity $D_{0}=v_0^2/(2D_r)$~\cite{Silke2011,Fily2012,XY2014}. 
After each time step, a new Voronoi tessellation is generated based on the updated cell positions. The cell shapes are determined in the process and the exchange of cell neighbors occurs naturally through topological  transitions~\cite{Bi2014}. 

We showed in Ref.~\cite{Bi2016} that the SPV model exhibits a transition from a solid-like state to a fluid-like state upon increasing  the single-cell motility $v_0$, the persistence time  $\tau_r$, or the cell shape parameter $P_0/\sqrt{A_0}$ that characterizes the competition between cell-cell adhesion and cortical tension. 
The phase diagram in the $(P_0,v_0)$ plane is reproduced in Fig.\ref{fig:spv2}B.  The transition is identified by setting the shape index $q=\langle P_i/\sqrt{A_i}\rangle$ to the value $q=3.813$, where $\langle...\rangle$ denotes the average over all cells. It was shown in Ref.~\cite{Bi2016} that the transition line located by $q=3.813$ coincides with the one based on the vanishing of the effective diffusivity obtained from the cellular mean-square-displacement.
Note that for fixed system size $A_{T}$, the preferred cell area $A_0$ does not affect the interaction forces or cellular shapes. Hence the solid-fluid  transition is insensitive to $A_0$, as demonstrated numerically 
and shown analytically in the SI.  The preferred area $A_0$ only shifts the total pressure of the tissue by a constant. 
\subsubsection*{Developing and validating traction-based mechanical inference}
\begin{figure}[!h]
\begin{centering}
\includegraphics[width=1.00\columnwidth]{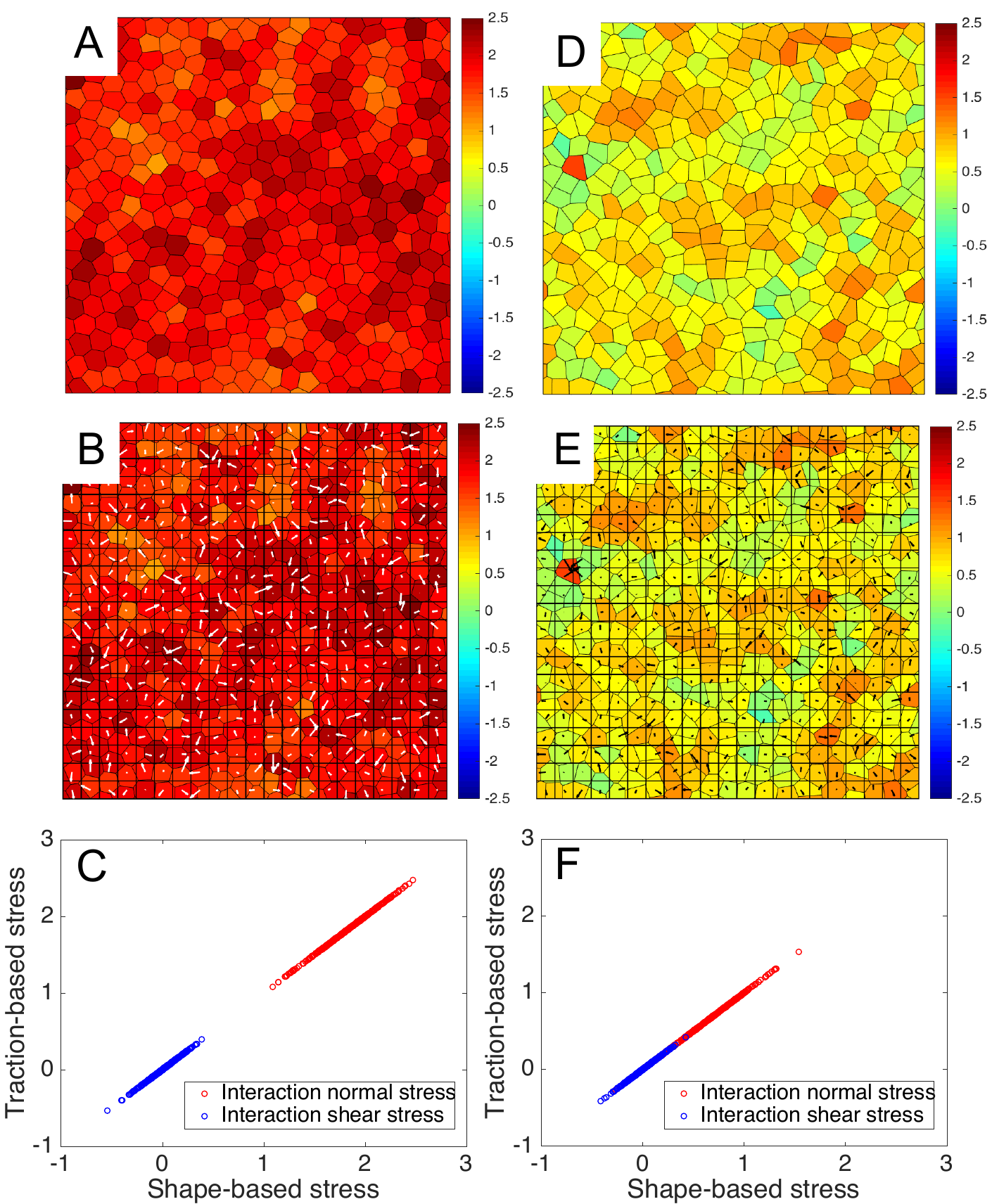}
\caption{Comparison of shape-based and coarse-grained traction-based stress. A-C: solid state at $v_0=0.5$, $P_0=3.3$. D-F: liquid state at $v_0=0.5$, $P_0=3.8$. A,D: Interaction normal stress $\sigma_n^{(i)int}$ calculated from the instantaneous cell shapes obtained from Eq.~\eqref{eqn:cellular_stress} and Eq.~\eqref{eqn:T_P}. Red denotes positive (contractile) stress and blue negative (extensile) stress. B,E: Interaction normal stress $\sigma_n^{(i)int}$ calculated using the coarse-grained traction-based mechanical inference by inverting ~Eq.~\eqref{force_balance_coarse} and using Eq.~\eqref{eqn:cellular_stress}. The arrows denote the traction forces. C,F: The coarse-grained traction-based mechanical inference is validated by plotting the traction-based stress against the shape-based stress in the solid (C) and in the liquid (F) state. The data are for $400$ cells in a square box of side $L=20$ with $D_r=0.1$ and with periodic boundary condition.}
\label{fig:distribution}
\end{centering}
\end{figure}
It is well established that in a vertex model tissue described by the tissue energy Eq.~\eqref{eqn:energy_square}, the mechanical state of cell $i$ is characterized by a local stress tensor $\sigma_{\alpha\beta}^{(i)int}$
given by~\cite{Ishihara2012,Jensen2016,Kevin2012}
\begin{subequations}
\begin{gather}
\label{eqn:cellular_stress}
\sigma_{\alpha\beta}^{(i)int}=-\Pi_i\delta_{\alpha\beta}+\frac{1}{2A_i}\sum_{ab\in i}T_{ab}^{\alpha}l_{ab}^{\beta},\\
\label{eqn:cellular_stress_b}
\Pi_i = -\frac{\partial E}{\partial A_i}\;,~~~~~~\mathbf{T}_{ab} = \frac{\partial E}{\partial \mathbf{l}_{ab}}\;,
\end{gather}
\end{subequations}
where $\Pi_i$ is the hydrostatic cellular pressure and $\mathbf{T}_{ab}=T_{ab}\mathbf{\hat{l}}_{ab}$ is the cell-edge tension, with $\mathbf{\hat{l}}_{ab}=\mathbf{l}_{ab}/|\mathbf{l}_{ab}|$. Here we use Roman indices $i,j,k,...$ to label cells and $a,b,c,...$ to label vertices, while Greek indices denote Cartesian components.  The summation in Eq.~\eqref{eqn:cellular_stress} runs over all edges of cell $i$,  $\mathbf{l}_{ab}$ is the edge vector joining vertices $a$ and $b$ 
when the perimeter of  cell $i$ is traversed clockwise, as shown in SI Fig.\ref{fig:inference}A. 
The factor of $1/2$ in the second term on the right hand side  of Eq.~\eqref{eqn:cellular_stress} is due to the fact that each edge is shared by two cells. We have used the convention that the cellular stress  is positive  when the cell is contractile and negative when the cell is extensile. Contractile stress means that if the cell is cut off from its neighbors, it tends to contract, consistent with the contractility of the actomyosin network within the cell. 

Note that in a vertex model the interaction stress as defined in  Eq.~\eqref{eqn:cellular_stress} is indeed the stress acting on the tissue boundary. This is, however, not the case for the Voronoi model because the Voronoi construction introduces constraints not present in the vertex model.  We have verified that the differences are small  (see SI text), and in the following use Eq.~\eqref{eqn:cellular_stress} as a good approximation for the Voronoi model.

Our goal is to obtain the distribution of cellular stresses in a layer of \textit{motile} cells, where cellular configurations do not minimize the tissue energy, but are governed by the dynamics described by Eq.~\eqref{equation_of_motion_SPV}. In this case, as discussed in the introduction, the local cellular stress can be written as the sum of contributions from interactions and propulsive forces as
\begin{equation}
\sigma_{\alpha\beta}^i=\sigma_{\alpha\beta}^{(i)int}+\sigma_{\alpha\beta}^{(i)swim}\;,
\label{sigma_i}
\end{equation}
where, following recent work on active colloids~\cite{XY2014,Takatori2014}, 
\begin{equation}
\sigma_{\alpha\beta}^{(i)swim}=-\frac{v_0}{\mu A_i}n_\alpha^ir_\beta^i
\label{swim_i}
\end{equation}
describes the flux of propulsive force $\frac{v_0}{\mu}\mathbf{n}_i$ across a boundary, as calculated from a virial expression. The negative sign in Eq.~\eqref{swim_i}  ensures that the swim stress follows the same convention as the interaction stress, i.e., is positive for contractile stress and negative for extensile stress. The swim stress  is proportional to the cell motility $v_0$ and vanishes for nonmotile cells. The contribution $\sigma_{\alpha\beta}^{(i)int}$ is still given by Eq.~\eqref{eqn:cellular_stress}, but depends implicitly on cell motility through the instantaneous values of $\Pi_i$ and $T_{ab}$ that are determined not by energy minimization, but by the system dynamics governed by Eq.~\eqref{equation_of_motion_SPV}. From the local stresses one can then obtain the total mean stress in the tissue as (see SI), $ \sigma_{\alpha\beta}=\sigma_{\alpha\beta}^{int} +\sigma_{\alpha\beta}^{swim}$, with
\begin{equation}
\sigma_{\alpha\beta}^{int} =\frac{1}{A_T}\sum_{i} A_i\sigma^{(i)int}_{\alpha\beta}\;,
\hspace{0.05in} \sigma_{\alpha\beta}^{swim}=\frac{1}{A_T}\sum_{i} A_i\sigma^{(i)swim}_{\alpha\beta}\;.
\label{tissue_level_stress}
\end{equation}

In simulations where the energy functional is known, it is simple to directly extract the instantaneous pressures and tensions from cell shapes in order to calculate the interaction contribution $\sigma_{\alpha\beta}^{(i)int}$. The definitions, 
 Eq.~\eqref{eqn:cellular_stress_b}, give
\begin{equation}
\label{eqn:T_P}
\Pi_i = -2K_A(A_i-A_0)\;,~T_{ab} = 2K_P[(P_j-P_0)+(P_k-P_0)]\;,
\end{equation}
where $ab$ is the cell-cell interface that separates cells $j$ and $k$. Both $A_i$ and $P_i$ are obtained at every time step of the simulation and implicitly depend on cell motility, as parametrized by speed $v_0$ and persistence $\tau_r$. This method directly infers the interaction stress from instantaneous cell shape fluctuations through Eq.~\eqref{eqn:cellular_stress}, yielding what we call \textit{shape-based stresses}. While easily implemented numerically, this method is of limited use in experiments where the energy functional is not known and likely more complicated. 

For this reason we develop a new mechanical inference method for motile monolayers that attempts to approximate the interaction stresses using only information that is accessible in experiments. Specifically, the proposed  {\em traction-based mechanical inference} infers tensions and pressures from segmented images of cell boundaries and traction forces obtained by TFM.
In the SPV model, we define the traction force at each vertex as the gradient of the tissue energy with respect to the vertex position $\mathbf{t}_a=-\bm\nabla_aE$, which balances the interaction force $\mathbf{F}_a$. Equilibrium mechanical inference methods express the interaction force $\mathbf{F}_a=-\bm\nabla_aE$ at each vertex in terms of cellular pressures and edge tensions and the measured geometry of the cellular network (see Eqs.~(\ref{decomposed}-\ref{TP2}) of SI).  Pressures and edge tensions are then obtained by inverting the equations 
$\mathbf{F}_a(\{\Pi_i\},\{T_{ab}\})=0$. For a nonequilibrium epithelial layer of motile cells we invert the force balance equations 
\begin{equation}
\mathbf{F}_a(\{\Pi_i\},\{T_{i}\})=\mathbf{t}_a\;,
\label{force_balance}
\end{equation}
where the edge tensions $T_{ab}$ have been written as the sum of cortical tensions $T_i$ of the adjacent cells (see Eq.~\eqref{TP1} of SI), reducing the number of independent unknowns. The interaction contribution to the local cellular stresses is then again calculated using Eq.~\eqref{eqn:cellular_stress}. 
A constraint counting yields $4N$ force balance equations for $2N$ variables, rendering the system overdetermined, which requires the implementation of a least square minimization for the mechanical inference (See SI Text).

The equations developed thus far require knowledge of the tractions at each vertex, which is again not realistic in experiments.
Therefore, we have developed and implemented a coarse-grained version of this approach that utilizes experimentally accessible traction forces averaged over a square grid, with a grid spacing of the order of a cell diameter. Pressures and tensions $(\Pi_i,T_i)$ are then calculated by inverting the force balance equation at the center of each grid element,
{\color{black}
\begin{equation}
\mathbf{F}_{grid}(\{\Pi_i\},\{T_{i}\})=\mathbf{t}_{grid}\;.
\label{force_balance_coarse}
\end{equation}
}
An outline of the coarse-graining procedure is given in the SI, with full details to be published elsewhere~\cite{XY2016}. We refer to stresses inferred from 
 Eq.~\eqref{force_balance_coarse} as \textit{traction-based stresses}. {\color{black}We emphasize that the traction-based stress from mechanical inference is generally different from the intercellular stress obtained with Monolayer Stress Microscopy (MSM), which rests upon the assumption that the tissue is an isotropic, homogeneous and linearly elastic material~\cite{Dhananjay2013}. The mechanical inference does not make such assumptions, and is compatible with any epithelia whose cell-cell interactions can be decomposed into tensions at cell junctions and pressures within cell bodies.} 

Within the framework of the SPV model, we have validated the coarse-grained method by comparing the resulting  \textit{traction-based} stresses to the \textit{shape-based} stresses computed exactly from the simulations. The two show perfect agreement (Fig.~\ref{fig:distribution}C,F). 

\subsubsection*{Stress characterizes rheological properties of the tissue}
To study the mechanical properties of motile confluent tissues, we simulate a confluent cell layer with periodic boundary conditions using the SPV model. By examining the temporal correlations of the mean stress in the tissue, as defined in Eq.~\eqref{tissue_level_stress}, we show that the tissue displays distinct mechanical properties in the liquid and in the solid states.  Thus mechanical measurement such as those provided by TFM can be used to characterize the rheological properties of the tissue.

 The stress tensor $\sigma_{\alpha\beta}$ is symmetric and has three independent components in $2d$. Both the mean and local stress components are most usefully expressed in terms of normal stress $\sigma_n$, shear stress $\sigma_s$ and normal stress difference $\sigma_d$, with
\begin{equation}
 \sigma_{n,d}=\frac{1}{2}(\sigma_{xx}\pm\sigma_{yy})\;,~~~\sigma_{s}=\frac{1}{2}(\sigma_{xy}+\sigma_{yx})\;.
 \label{eqn:stress_component}
 \end{equation}
Each of the interaction and swim contributions can similarly be split in normal, shear and normal difference components.  Below we focus on normal and shear stresses. 
TFM probes the forces exchanged between tissue and substrate, which by force balance are determined entirely by intercellular forces, hence by interaction stresses.  
 In contrast, the swim components of stress and pressure cannot be probed in TFM, but contribute to  the pressure $\Pi=-\sigma_n=\Pi_{int}+\Pi_{swim}$ that the tissue would exert laterally on a confining piston. As we will see below, the swim contribution dominates the pressure in the liquid state.
 
Using the expression for the local stress obtained from cell shapes, the mean interaction normal stress of the tissue can be expressed entirely in terms of area and perimeter fluctuations in a virial-like form (See SI Text)
\begin{equation}
\sigma_n^{int} = \frac{1}{A_T}\sum_i\left[2K_A A_i(A_i-A_0) +K_P P_i(P_i-P_0)\right]\;.
\label{eqn:tissue_normal_main}
\end{equation}
The first term represents the interaction contribution from the pressures within the cells. The second term is the contribution from the competition between actomyosin contractility and cell-cell adhesion that controls the cortical tensions. In our simulation, the cellular pressure is suppressed by setting $\bar{A}=A_0$ and the normal stress comes mainly from the cortical tensions. Equation~\ref{eqn:tissue_normal_main} then provides a way for extracting mechanical information directly from cell shape based on snapshots of segmented cell images.

\textit{Normal stresses are contractile in the solid phase and become extensile deep in the liquid phase.}
We show in Fig.~\ref{fig:distribution} snapshots of the local interaction normal stress in the solid state (left column) and in the liquid state (right column).
In both the solid and the liquid states the interaction normal stress is on average contractile (red), with relatively weaker spatial fluctuations, but much larger mean value in the solid state, where contractile cortical tension exceeds cell-cell adhesion. 
\begin{figure}[!h]
\begin{centering}
\includegraphics[width=1.00\columnwidth]{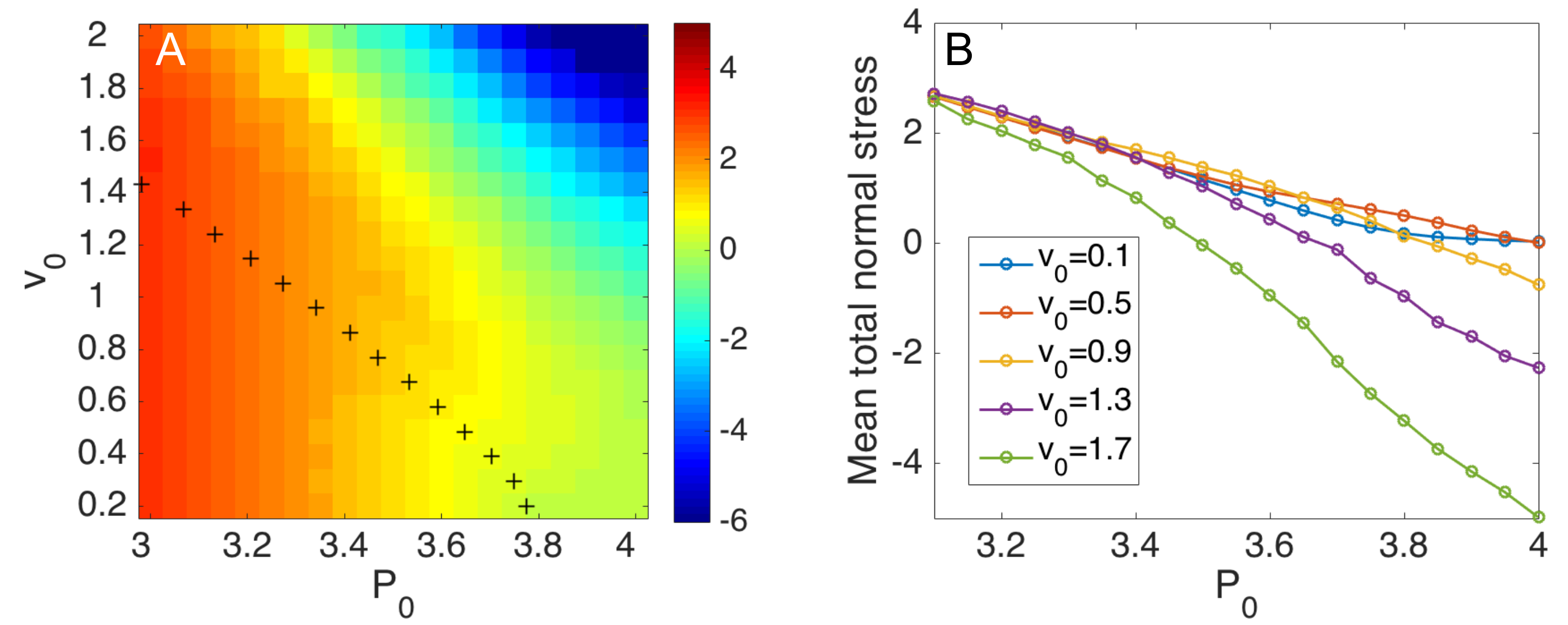}
\caption{Mean total normal stress. A: heat map of the mean total normal stress of the tissue in the $(v_0,P_0)$ plane.  The black crosses outline the solid-liquid phase boundary determined by $q=3.813$. Red indicates contractile stress and blue extensile stress. B: mean total normal stress as a function of $P_0$ at various $v_0$, showing a change in sign deep in the liquid state. ($400$ cells for $T=1000$ and $D_r=0.1$ with periodic boundary condition)}
\label{fig:normal_stress}
\end{centering}
\end{figure}
Fig.~\ref{fig:normal_stress} displays the total mean normal stress (the separate contributions from interaction and swim stress are shown in {\color{black} SI Fig.\ref{fig:int_swim}}) across the solid-liquid transition.
The color map shows that the total normal stress is contractile in the solid state and across the transition line denoted by the black crosses, but changes sign and becomes extensile deep in the liquid state. While the interaction stress is always positive due to cell contractility and consistent with experimental observations~\cite{Lecuit2007,Kapil2014,Gardel2015,Bellaiche2013,Rauzi2010,Zallen2014}, the change in sign of the total stress is due to the swim stress that is zero in the solid and always negative in the liquid (see {\color{black} SI Fig.\ref{fig:int_swim}}), indicating that motility induces extensile stresses tending to stretch the tissue.
 The total normal stress is analogous to the stress on a wall confining an active Brownian colloidal fluid~\cite{XY2014,Takatori2014}.  We speculate that its change in sign could lead to an expansion of the tissue if released from confinement due to substrate patterning or to surrounding tissue, and may contribute to epithelia expansion in wound healing assays. In our model confinement is  provided  by the periodic boundary conditions.

\begin{figure}
\begin{centering}
\includegraphics[width=1.00\columnwidth]{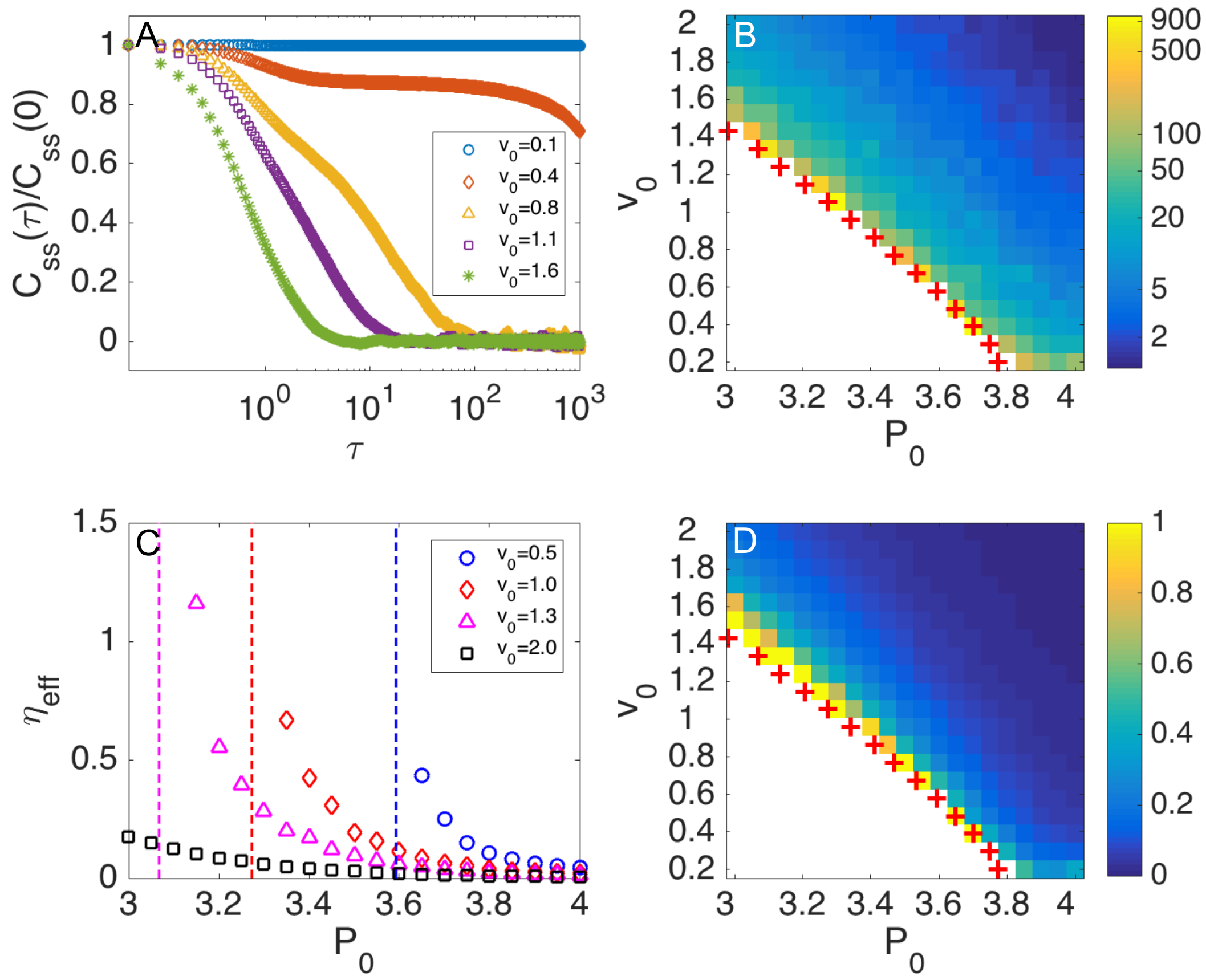}
\caption{Time autocorrelation of interaction shear stress and effective tissue viscosity. A: Time autocorrelation function of the mean interaction shear stress for  $P_0=3.45$ at various $v_0$. B and D: Heat maps of the correlation time $\tau_m$ (B) and of the effective viscosity (D). The red crosses denote the liquid-solid phase transition boundary determined by $q=3.813$. The white region corresponds to regions where correlation time $\tau_m$ and effective viscosity $\eta_\mathit{eff}$ diverge. Note that these regions largely coincide with the solid regime.
C: The effective viscosity as a function of $P_0$ at various $v_0$. The dashed lines correspond to the critical $P_0$ where the liquid-solid phase transition occurs. All results are for $100$ cells, $D_r=1$ and $T=40000$ with periodic boundary condition.}
\label{fig:viscosity}
\end{centering}
\end{figure}

\textit{The tissue effective shear viscosity diverges at the liquid-solid transition.} 
While the local shear stress averages to zero in both the liquid and the solid states, its temporal correlations provide a distinctive rheological metric for distinguishing the liquid from the solid and identifying the transition.  
The time autocorrelation function of the interaction shear stress,
\begin{equation}
\label{eqn:corr}
C_{ss}(\tau) = \langle \sigma_{s}^{int}(t_0)\sigma_{s}^{int}(t_0+\tau)\rangle_{t_0}\;,
\end{equation}
where $\langle...\rangle_{t_0}$ denotes the average over the length $t_0$ of the simulation, is shown in Fig.~\ref{fig:viscosity}A for various $v_0$ across the liquid-solid transition. Shear stress correlations decay in the liquid state and slow down as the transition is approached from the liquid side, ultimately becoming frozen in the solid state. To quantify this we have defined the correlation time $\tau_m$  as the time when the correlation has decreased below one percent of its initial value. This correlation time shown in Fig.~\ref{fig:viscosity}B diverges at the liquid-solid phase transition, suggesting
that shear stress autocorrelations can provide a robust metric for the transition.  
Our work therefore suggests that TFM {\color{black}combined with mechanical inference} can provide a tool for the measurement of tissue rheology. Correlating such measurements with cell shape data will provide a stringent test for our theory.  Finally, in the liquid state we  define an effective viscosity $\eta_{\mathit{eff}}$ using a Green-Kubo-type relation by integrating the correlation function over the duration of the correlation time~\cite{Hansen2005,Egami2011},
\begin{equation}
\label{eqn:green_kubo}
\eta_{\mathit{eff}} = \frac{A_T}{k_bT_{\mathit{eff}}}\int_{0}^{\tau_m}C_{ss}(\tau)d\tau\;,
\end{equation}
where we have used the ideal gas effective temperature $k_B T_{\mathit{eff}}=v_0^2/(2\mu D_r)$ so that $\eta_{\mathit{eff}}$ has dimensions of a shear viscosity in $2d$. Of course the Green-Kubo relation is based on the existence of a fluctuation-dissipation theorem, which does not hold in active systems. For small values of the persistence time $\tau_r$, however, the orientational noise in the SPV becomes uncorrelated in time  and can be mapped onto thermal noise at an effective temperature $T_{\mathit{eff}}$. In this limit we expect $\eta_{\mathit{eff}}$ to indeed play the role of a shear viscosity. 
Remarkably, the effective viscosity shown in Fig.\ref{fig:viscosity}(C,D) grows as the tissue approaches the solid state from the liquid side, and diverges at the transition. The effective viscosity quickly approaches zero deep in the liquid phase, suggesting that the system behaves as a gas of uncorrelated cells in this region.
\\
\section*{Discussion and Conclusions}
Using the Self-propelled Voronoi Model (SPV) we have formulated a unifying framework for quantifying the contributions from cell shape fluctuations and cell motility to mechanical stresses in an epithelial tissue.   Cell shape fluctuations  from actomyosin contractility and cell-cell adhesion control the interaction stress, while cell motility determines the swim stress that is generically present in all self-propelled systems~\cite{XY2014,Takatori2014}.

Unlike monolayer stress microscopy that computes interaction stress from traction forces by assuming the tissue to be a continuum linear elastic material~\cite{Dhananjay2013}, the \textit{traction-based mechanical inference} method developed here incorporates spatial and temporal deformations of the tissue due to actomyosin contractility and cell-cell adhesion, and can be generalized to account for cell division and apoptosis. In contrast to equilibrium mechanical inference techniques~\cite{Kevin2012,Ishihara2012}, our approach does not require cells to be in or close to static mechanical balance, and it also provides the absolute scale of the junctional tensions and pressure differences. This can for instance be important for testing hypotheses involving mechanosensitive biomolecules. Experimentally, our method provides new ways to extract inter-cellular interaction stresses from existing traction force data and segmented cell images. 

The swim stress, on the other hand, cannot be measured using TFM as it represents the flux of propulsive forces across a bulk plane in the tissue. It contributes to the homeostatic pressure at the lateral boundary of the tissue. The sum of the swim stress and the interaction stress approximates the total stress at the tissue boundary, which is generally contractile but can become extensile when the tissue is deep in the liquid state and cell motility exceeds actomyosin contractility. This change in sign may be observable in wound healing assays where the transition from contractile  to extensile behavior can result in tissue expansion upon removal of  confinement by  neighboring tissue. The application of our method to the wound healing geometry will be described in future work.

We have extracted an effective tissue viscosity from the temporal correlation of the interaction shear stress. The correlation time and effective viscosity display a slowing down and arrest at the transition to the solid regime, thus serve as a direct probe of tissue rheology.  Moreover, we observed a similar behavior for the temporal correlations of traction forces as demonstrated in the SI. Therefore, our work suggests that TFM measurement~\cite{Dhananjay2011} combined with mechanical inference could provide information on tissue rheology. To our knowledge, this has not been attempted yet on experimental data. 

Our work sets the stage for examining the feedback between cell activity and tissue mechanics that is apparent in many tissue-level phenomena. Recent work has shown that mechanical stresses influence cell proliferation in tumor spheroids~\cite{Montel2011} and regulate cell growth in the developing \textit{Drosophila} wing~\cite{Pan2016}. Regulation of cell  motility, as in contact inhibition of locomotion, has been proposed to explain stress patterns during collective cell migration~\cite{Zimmermann2016}. TFM has revealed the tendency of cells to move along the direction of minimal shear stress, a phenomenon termed ``plithotaxis''~\cite{Dhananjay2011}. Our model provides a unifying framework for quantifying the relative roles of various cell properties, such as shape as controlled by contractility and cell-cell adhesion, motility and growth, on the mechanics of the tissue.

We thank Daniel Sussman for valuable discussions. This work was supported by the Simons Foundation through a Targeted Grant Award No. 342354 (MCM \& MC) and an Investigator Award No. 446222 (MLM, MC, \& MM) in the Mathematical Modeling of Living Systems, by the National Science Foundation through awards DMR-1305184 (MCM \& XY), DMR- 1609208 (MCM), DMR-1352184 (MLM \& DB), and the IGERT grant DGE-1068780 (MCM \& MC), the NIH through R01GM117598-02 (MLM), and by the computational resources provided by Syracuse University and through NSF ACI-1541396. All authors acknowledge support from the Syracuse University Soft Matter Program.

\bibliography{References.bib}

\begin{thebibliography}{43}
\expandafter\ifx\csname natexlab\endcsname\relax\def\natexlab#1{#1}\fi
\expandafter\ifx\csname bibnamefont\endcsname\relax
  \def\bibnamefont#1{#1}\fi
\expandafter\ifx\csname bibfnamefont\endcsname\relax
  \def\bibfnamefont#1{#1}\fi
\expandafter\ifx\csname citenamefont\endcsname\relax
  \def\citenamefont#1{#1}\fi
\expandafter\ifx\csname url\endcsname\relax
  \def\url#1{\texttt{#1}}\fi
\expandafter\ifx\csname urlprefix\endcsname\relax\def\urlprefix{URL }\fi
\providecommand{\bibinfo}[2]{#2}
\providecommand{\eprint}[2][]{\url{#2}}

\bibitem[{\citenamefont{Lecuit and Lenne}(2007)}]{Lecuit2007}
\bibinfo{author}{\bibfnamefont{T.}~\bibnamefont{Lecuit}} \bibnamefont{and}
  \bibinfo{author}{\bibfnamefont{P.-F.} \bibnamefont{Lenne}},
  \bibinfo{journal}{nature molecular cell biology} \textbf{\bibinfo{volume}{8}}
  (\bibinfo{year}{2007}).

\bibitem[{\citenamefont{Heisenberg and Bellaiche}(2013)}]{Bellaiche2013}
\bibinfo{author}{\bibfnamefont{C.-P.} \bibnamefont{Heisenberg}}
  \bibnamefont{and}
  \bibinfo{author}{\bibfnamefont{Y.}~\bibnamefont{Bellaiche}},
  \bibinfo{journal}{Cell} pp. \bibinfo{pages}{Volume 153, Issue 5, p948--962}
  (\bibinfo{year}{2013}).

\bibitem[{\citenamefont{Murrell et~al.}(2015)\citenamefont{Murrell, Oakes,
  Lenz, and Gardel}}]{Gardel2015}
\bibinfo{author}{\bibfnamefont{M.}~\bibnamefont{Murrell}},
  \bibinfo{author}{\bibfnamefont{P.~W.} \bibnamefont{Oakes}},
  \bibinfo{author}{\bibfnamefont{M.}~\bibnamefont{Lenz}}, \bibnamefont{and}
  \bibinfo{author}{\bibfnamefont{M.~L.} \bibnamefont{Gardel}},
  \bibinfo{journal}{Nat Rev Mol Cell Biol.} pp. \bibinfo{pages}{16(8):486--98}
  (\bibinfo{year}{2015}).

\bibitem[{\citenamefont{Etournay et~al.}(2015)\citenamefont{Etournay, Popovic,
  Merkel, Nandi, Blasse, Aigouy, Brandl, Myers, Salbreux, Julicher
  et~al.}}]{Eaton2015}
\bibinfo{author}{\bibfnamefont{R.}~\bibnamefont{Etournay}},
  \bibinfo{author}{\bibfnamefont{M.}~\bibnamefont{Popovic}},
  \bibinfo{author}{\bibfnamefont{M.}~\bibnamefont{Merkel}},
  \bibinfo{author}{\bibfnamefont{A.}~\bibnamefont{Nandi}},
  \bibinfo{author}{\bibfnamefont{C.}~\bibnamefont{Blasse}},
  \bibinfo{author}{\bibfnamefont{B.}~\bibnamefont{Aigouy}},
  \bibinfo{author}{\bibfnamefont{H.}~\bibnamefont{Brandl}},
  \bibinfo{author}{\bibfnamefont{G.}~\bibnamefont{Myers}},
  \bibinfo{author}{\bibfnamefont{G.}~\bibnamefont{Salbreux}},
  \bibinfo{author}{\bibfnamefont{F.}~\bibnamefont{Julicher}},
  \bibnamefont{et~al.}, \bibinfo{journal}{eLife} \textbf{\bibinfo{volume}{eLife
  2015;4:e07090}} (\bibinfo{year}{2015}).

\bibitem[{\citenamefont{Shraiman}(2005)}]{Shraiman2005}
\bibinfo{author}{\bibfnamefont{B.~I.} \bibnamefont{Shraiman}},
  \bibinfo{journal}{PNAS} \textbf{\bibinfo{volume}{102}}, \bibinfo{pages}{3318}
  (\bibinfo{year}{2005}).

\bibitem[{\citenamefont{Serra-Picamal et~al.}(2012)\citenamefont{Serra-Picamal,
  Conte, Vincent, Anon, Tambe, Bazellieres, Butler, Fredberg, and
  Trepat}}]{Xavier2012}
\bibinfo{author}{\bibfnamefont{X.}~\bibnamefont{Serra-Picamal}},
  \bibinfo{author}{\bibfnamefont{V.}~\bibnamefont{Conte}},
  \bibinfo{author}{\bibfnamefont{R.}~\bibnamefont{Vincent}},
  \bibinfo{author}{\bibfnamefont{E.}~\bibnamefont{Anon}},
  \bibinfo{author}{\bibfnamefont{D.~T.} \bibnamefont{Tambe}},
  \bibinfo{author}{\bibfnamefont{E.}~\bibnamefont{Bazellieres}},
  \bibinfo{author}{\bibfnamefont{J.~P.} \bibnamefont{Butler}},
  \bibinfo{author}{\bibfnamefont{J.~J.} \bibnamefont{Fredberg}},
  \bibnamefont{and} \bibinfo{author}{\bibfnamefont{X.}~\bibnamefont{Trepat}},
  \bibinfo{journal}{Nat. Phys.} pp. \bibinfo{pages}{8, 628--634}
  (\bibinfo{year}{2012}).

\bibitem[{\citenamefont{Staple et~al.}(2010)\citenamefont{Staple, Farhadifar,
  R{\"{o}}per, Aigouy, Eaton, and J{\"{u}}licher}}]{Staple2010}
\bibinfo{author}{\bibfnamefont{D.~B.} \bibnamefont{Staple}},
  \bibinfo{author}{\bibfnamefont{R.}~\bibnamefont{Farhadifar}},
  \bibinfo{author}{\bibfnamefont{J.-C.} \bibnamefont{R{\"{o}}per}},
  \bibinfo{author}{\bibfnamefont{B.}~\bibnamefont{Aigouy}},
  \bibinfo{author}{\bibfnamefont{S.}~\bibnamefont{Eaton}}, \bibnamefont{and}
  \bibinfo{author}{\bibfnamefont{F.}~\bibnamefont{J{\"{u}}licher}},
  \bibinfo{journal}{The European physical journal. E, Soft matter}
  \textbf{\bibinfo{volume}{33}}, \bibinfo{pages}{117} (\bibinfo{year}{2010}),
  ISSN \bibinfo{issn}{1292-895X},
  \urlprefix\url{http://www.ncbi.nlm.nih.gov/pubmed/21082210}.

\bibitem[{\citenamefont{Guillot and Lecuit}(2013)}]{Guillot2013}
\bibinfo{author}{\bibfnamefont{C.}~\bibnamefont{Guillot}} \bibnamefont{and}
  \bibinfo{author}{\bibfnamefont{T.}~\bibnamefont{Lecuit}},
  \bibinfo{journal}{Science} \textbf{\bibinfo{volume}{340}},
  \bibinfo{pages}{1185} (\bibinfo{year}{2013}).

\bibitem[{\citenamefont{Rauzzi et~al.}(2008)\citenamefont{Rauzzi, Verant,
  Lecuit, and Lenne}}]{Lenne2008}
\bibinfo{author}{\bibfnamefont{M.}~\bibnamefont{Rauzzi}},
  \bibinfo{author}{\bibfnamefont{P.}~\bibnamefont{Verant}},
  \bibinfo{author}{\bibfnamefont{T.}~\bibnamefont{Lecuit}}, \bibnamefont{and}
  \bibinfo{author}{\bibfnamefont{P.-F.} \bibnamefont{Lenne}},
  \bibinfo{journal}{nature cell biology} \textbf{\bibinfo{volume}{10}}
  (\bibinfo{year}{2008}).

\bibitem[{\citenamefont{Trepat et~al.}(2009)\citenamefont{Trepat, R.Wasserman,
  Angelini, Millet, A.Weitz, Butler, and Fredberg}}]{Xavier2009}
\bibinfo{author}{\bibfnamefont{X.}~\bibnamefont{Trepat}},
  \bibinfo{author}{\bibfnamefont{M.}~\bibnamefont{R.Wasserman}},
  \bibinfo{author}{\bibfnamefont{T.~E.} \bibnamefont{Angelini}},
  \bibinfo{author}{\bibfnamefont{E.}~\bibnamefont{Millet}},
  \bibinfo{author}{\bibfnamefont{D.}~\bibnamefont{A.Weitz}},
  \bibinfo{author}{\bibfnamefont{J.~P.} \bibnamefont{Butler}},
  \bibnamefont{and} \bibinfo{author}{\bibfnamefont{J.~J.}
  \bibnamefont{Fredberg}}, \bibinfo{journal}{Nat. Phys.} pp. \bibinfo{pages}{5,
  426 -- 430} (\bibinfo{year}{2009}).

\bibitem[{\citenamefont{Tambe et~al.}(2011)\citenamefont{Tambe, Hardin,
  Angelini, Rajendran, Park, Serra-Picamal, Zhou, Zaman, Butler, A.Weitz
  et~al.}}]{Dhananjay2011}
\bibinfo{author}{\bibfnamefont{D.~T.} \bibnamefont{Tambe}},
  \bibinfo{author}{\bibfnamefont{C.~C.} \bibnamefont{Hardin}},
  \bibinfo{author}{\bibfnamefont{T.~E.} \bibnamefont{Angelini}},
  \bibinfo{author}{\bibfnamefont{K.}~\bibnamefont{Rajendran}},
  \bibinfo{author}{\bibfnamefont{C.~Y.} \bibnamefont{Park}},
  \bibinfo{author}{\bibfnamefont{X.}~\bibnamefont{Serra-Picamal}},
  \bibinfo{author}{\bibfnamefont{E.~H.} \bibnamefont{Zhou}},
  \bibinfo{author}{\bibfnamefont{M.~H.} \bibnamefont{Zaman}},
  \bibinfo{author}{\bibfnamefont{J.~P.} \bibnamefont{Butler}},
  \bibinfo{author}{\bibfnamefont{D.}~\bibnamefont{A.Weitz}},
  \bibnamefont{et~al.}, \bibinfo{journal}{Nat. Mater.} pp.
  \bibinfo{pages}{10(6):469--75} (\bibinfo{year}{2011}).

\bibitem[{\citenamefont{Bi et~al.}(2014)\citenamefont{Bi, Lopez, Schwarz, and
  Manning}}]{Bi2014}
\bibinfo{author}{\bibfnamefont{D.}~\bibnamefont{Bi}},
  \bibinfo{author}{\bibfnamefont{J.~H.} \bibnamefont{Lopez}},
  \bibinfo{author}{\bibfnamefont{J.~M.} \bibnamefont{Schwarz}},
  \bibnamefont{and} \bibinfo{author}{\bibfnamefont{M.~L.}
  \bibnamefont{Manning}}, \bibinfo{journal}{Soft Matter} pp.
  \bibinfo{pages}{10, 1885} (\bibinfo{year}{2014}).

\bibitem[{\citenamefont{Bi et~al.}(2015)\citenamefont{Bi, Lopez, Schwarz, and
  Manning}}]{Manning2015}
\bibinfo{author}{\bibfnamefont{D.}~\bibnamefont{Bi}},
  \bibinfo{author}{\bibfnamefont{J.~H.} \bibnamefont{Lopez}},
  \bibinfo{author}{\bibfnamefont{J.~M.} \bibnamefont{Schwarz}},
  \bibnamefont{and} \bibinfo{author}{\bibfnamefont{M.~L.}
  \bibnamefont{Manning}}, \bibinfo{journal}{Nat. Phys.}
  \textbf{\bibinfo{volume}{11}} (\bibinfo{year}{2015}).

\bibitem[{\citenamefont{Bi et~al.}(2016)\citenamefont{Bi, Yang, Marchetti, and
  Manning}}]{Bi2016}
\bibinfo{author}{\bibfnamefont{D.}~\bibnamefont{Bi}},
  \bibinfo{author}{\bibfnamefont{X.}~\bibnamefont{Yang}},
  \bibinfo{author}{\bibfnamefont{M.~C.} \bibnamefont{Marchetti}},
  \bibnamefont{and} \bibinfo{author}{\bibfnamefont{M.~L.}
  \bibnamefont{Manning}}, \bibinfo{journal}{Phys. Rev. X}
  \textbf{\bibinfo{volume}{6}} (\bibinfo{year}{2016}).

\bibitem[{\citenamefont{Park et~al.}(2015)\citenamefont{Park, Kim, Bi, Mitchel,
  Qazvini, Tantisira, Park, McGill, Kim, Gweon et~al.}}]{Park2015}
\bibinfo{author}{\bibfnamefont{J.-A.} \bibnamefont{Park}},
  \bibinfo{author}{\bibfnamefont{J.~H.} \bibnamefont{Kim}},
  \bibinfo{author}{\bibfnamefont{D.}~\bibnamefont{Bi}},
  \bibinfo{author}{\bibfnamefont{J.~A.} \bibnamefont{Mitchel}},
  \bibinfo{author}{\bibfnamefont{N.~T.} \bibnamefont{Qazvini}},
  \bibinfo{author}{\bibfnamefont{K.}~\bibnamefont{Tantisira}},
  \bibinfo{author}{\bibfnamefont{C.~Y.} \bibnamefont{Park}},
  \bibinfo{author}{\bibfnamefont{M.}~\bibnamefont{McGill}},
  \bibinfo{author}{\bibfnamefont{S.-H.} \bibnamefont{Kim}},
  \bibinfo{author}{\bibfnamefont{B.}~\bibnamefont{Gweon}},
  \bibnamefont{et~al.}, \bibinfo{journal}{Nat. Mater.} pp. \bibinfo{pages}{14,
  1040--1048} (\bibinfo{year}{2015}).

\bibitem[{\citenamefont{Soin{\'e} et~al.}(2014)\citenamefont{Soin{\'e}, Brand,
  Stricker, Oakes, Gardel, and Schwarz}}]{Gardel2014}
\bibinfo{author}{\bibfnamefont{J.~R.~D.} \bibnamefont{Soin{\'e}}},
  \bibinfo{author}{\bibfnamefont{C.~A.} \bibnamefont{Brand}},
  \bibinfo{author}{\bibfnamefont{J.}~\bibnamefont{Stricker}},
  \bibinfo{author}{\bibfnamefont{P.~W.} \bibnamefont{Oakes}},
  \bibinfo{author}{\bibfnamefont{M.~L.} \bibnamefont{Gardel}},
  \bibnamefont{and} \bibinfo{author}{\bibfnamefont{U.~S.}
  \bibnamefont{Schwarz}}, \bibinfo{journal}{Plos Comp. Biol.}
  \textbf{\bibinfo{volume}{11(3):e1004076}} (\bibinfo{year}{2014}).

\bibitem[{\citenamefont{Tambe et~al.}(2013)\citenamefont{Tambe, Croutelle,
  Trepat, Park, Kim, Millet, Butler, and Fredberg}}]{Dhananjay2013}
\bibinfo{author}{\bibfnamefont{D.~T.} \bibnamefont{Tambe}},
  \bibinfo{author}{\bibfnamefont{U.}~\bibnamefont{Croutelle}},
  \bibinfo{author}{\bibfnamefont{X.}~\bibnamefont{Trepat}},
  \bibinfo{author}{\bibfnamefont{C.~Y.} \bibnamefont{Park}},
  \bibinfo{author}{\bibfnamefont{J.~H.} \bibnamefont{Kim}},
  \bibinfo{author}{\bibfnamefont{E.}~\bibnamefont{Millet}},
  \bibinfo{author}{\bibfnamefont{J.~P.} \bibnamefont{Butler}},
  \bibnamefont{and} \bibinfo{author}{\bibfnamefont{J.~J.}
  \bibnamefont{Fredberg}}, \bibinfo{journal}{Plos one} p. \bibinfo{pages}{8(2):
  e55172} (\bibinfo{year}{2013}).

\bibitem[{\citenamefont{Zimmermann et~al.}(2014)\citenamefont{Zimmermann,
  Hayes, Basan, Onuchic, Rappel, and Levine}}]{Zimmerman2014}
\bibinfo{author}{\bibfnamefont{J.}~\bibnamefont{Zimmermann}},
  \bibinfo{author}{\bibfnamefont{R.~L.} \bibnamefont{Hayes}},
  \bibinfo{author}{\bibfnamefont{M.}~\bibnamefont{Basan}},
  \bibinfo{author}{\bibfnamefont{J.~N.} \bibnamefont{Onuchic}},
  \bibinfo{author}{\bibfnamefont{W.-J.} \bibnamefont{Rappel}},
  \bibnamefont{and} \bibinfo{author}{\bibfnamefont{H.}~\bibnamefont{Levine}},
  \bibinfo{journal}{Biophysical journal} \textbf{\bibinfo{volume}{107(3)}}
  (\bibinfo{year}{2014}).

\bibitem[{\citenamefont{Kim et~al.}(2013)\citenamefont{Kim, Serra-Picamal,
  Tambe, Zhou, Park, Sadati, Park, Krishnan, Gweon, Butler et~al.}}]{Jae2013}
\bibinfo{author}{\bibfnamefont{J.~H.} \bibnamefont{Kim}},
  \bibinfo{author}{\bibfnamefont{X.}~\bibnamefont{Serra-Picamal}},
  \bibinfo{author}{\bibfnamefont{D.~T.} \bibnamefont{Tambe}},
  \bibinfo{author}{\bibfnamefont{E.~H.} \bibnamefont{Zhou}},
  \bibinfo{author}{\bibfnamefont{C.~Y.} \bibnamefont{Park}},
  \bibinfo{author}{\bibfnamefont{M.}~\bibnamefont{Sadati}},
  \bibinfo{author}{\bibfnamefont{J.-A.} \bibnamefont{Park}},
  \bibinfo{author}{\bibfnamefont{R.}~\bibnamefont{Krishnan}},
  \bibinfo{author}{\bibfnamefont{B.}~\bibnamefont{Gweon}},
  \bibinfo{author}{\bibfnamefont{E.~M. J.~P.} \bibnamefont{Butler}},
  \bibnamefont{et~al.}, \bibinfo{journal}{Nat. Mater.} pp. \bibinfo{pages}{12,
  856--863} (\bibinfo{year}{2013}).

\bibitem[{\citenamefont{Ishihara and Sugimura}(2012)}]{Ishihara2012}
\bibinfo{author}{\bibfnamefont{S.}~\bibnamefont{Ishihara}} \bibnamefont{and}
  \bibinfo{author}{\bibfnamefont{K.}~\bibnamefont{Sugimura}},
  \bibinfo{journal}{Journal of Theoretical Biology} pp. \bibinfo{pages}{313,
  201--211} (\bibinfo{year}{2012}).

\bibitem[{\citenamefont{Nier et~al.}(2016)\citenamefont{Nier, Jain, Lim,
  Ishihara, Ladoux, and Marcq}}]{Marcp2016}
\bibinfo{author}{\bibfnamefont{V.}~\bibnamefont{Nier}},
  \bibinfo{author}{\bibfnamefont{S.}~\bibnamefont{Jain}},
  \bibinfo{author}{\bibfnamefont{C.~T.} \bibnamefont{Lim}},
  \bibinfo{author}{\bibfnamefont{S.}~\bibnamefont{Ishihara}},
  \bibinfo{author}{\bibfnamefont{B.}~\bibnamefont{Ladoux}}, \bibnamefont{and}
  \bibinfo{author}{\bibfnamefont{P.}~\bibnamefont{Marcq}},
  \bibinfo{journal}{Biophysical journal} \textbf{\bibinfo{volume}{110}}
  (\bibinfo{year}{2016}).

\bibitem[{\citenamefont{Chiou et~al.}(2012)\citenamefont{Chiou, Hufnagel, and
  Shraiman}}]{Kevin2012}
\bibinfo{author}{\bibfnamefont{K.~K.} \bibnamefont{Chiou}},
  \bibinfo{author}{\bibfnamefont{L.}~\bibnamefont{Hufnagel}}, \bibnamefont{and}
  \bibinfo{author}{\bibfnamefont{B.~I.} \bibnamefont{Shraiman}},
  \bibinfo{journal}{Plos Comp. Biol.} p. \bibinfo{pages}{8(5): e1002512}
  (\bibinfo{year}{2012}).

\bibitem[{\citenamefont{Noll et~al.}(2015)\citenamefont{Noll, Mani, Heemskerk,
  Streichan, and Shraiman}}]{Noll2015}
\bibinfo{author}{\bibfnamefont{N.}~\bibnamefont{Noll}},
  \bibinfo{author}{\bibfnamefont{M.}~\bibnamefont{Mani}},
  \bibinfo{author}{\bibfnamefont{I.}~\bibnamefont{Heemskerk}},
  \bibinfo{author}{\bibfnamefont{S.}~\bibnamefont{Streichan}},
  \bibnamefont{and} \bibinfo{author}{\bibfnamefont{B.~I.}
  \bibnamefont{Shraiman}}, \bibinfo{journal}{arXiv:1508.00623}
  (\bibinfo{year}{2015}).

\bibitem[{\citenamefont{Brodland et~al.}(2010)\citenamefont{Brodland, Conte,
  Cranston, Veldhuis, Narasimhan, Hutson, Jacinto, Ulrich, Baum, and
  Miodownik}}]{Brodland2010}
\bibinfo{author}{\bibfnamefont{G.~W.} \bibnamefont{Brodland}},
  \bibinfo{author}{\bibfnamefont{V.}~\bibnamefont{Conte}},
  \bibinfo{author}{\bibfnamefont{P.~G.} \bibnamefont{Cranston}},
  \bibinfo{author}{\bibfnamefont{J.}~\bibnamefont{Veldhuis}},
  \bibinfo{author}{\bibfnamefont{S.}~\bibnamefont{Narasimhan}},
  \bibinfo{author}{\bibfnamefont{M.~S.} \bibnamefont{Hutson}},
  \bibinfo{author}{\bibfnamefont{A.}~\bibnamefont{Jacinto}},
  \bibinfo{author}{\bibfnamefont{F.}~\bibnamefont{Ulrich}},
  \bibinfo{author}{\bibfnamefont{B.}~\bibnamefont{Baum}}, \bibnamefont{and}
  \bibinfo{author}{\bibfnamefont{M.}~\bibnamefont{Miodownik}},
  \bibinfo{journal}{PNAS} pp. \bibinfo{pages}{22111--22116}
  (\bibinfo{year}{2010}).

\bibitem[{\citenamefont{Farhadifar et~al.}(2007)\citenamefont{Farhadifar,
  Roper, Aigouy, Eaton, and Julicher}}]{Farhadifar2007}
\bibinfo{author}{\bibfnamefont{R.}~\bibnamefont{Farhadifar}},
  \bibinfo{author}{\bibfnamefont{J.-C.} \bibnamefont{Roper}},
  \bibinfo{author}{\bibfnamefont{B.}~\bibnamefont{Aigouy}},
  \bibinfo{author}{\bibfnamefont{S.}~\bibnamefont{Eaton}}, \bibnamefont{and}
  \bibinfo{author}{\bibfnamefont{F.}~\bibnamefont{Julicher}},
  \bibinfo{journal}{Current Biology} pp. \bibinfo{pages}{17, 2095--2104}
  (\bibinfo{year}{2007}).

\bibitem[{\citenamefont{Basan et~al.}(2009)\citenamefont{Basan, Risler, Joanny,
  Sastre-Garau, and Prost}}]{Prost2009}
\bibinfo{author}{\bibfnamefont{M.}~\bibnamefont{Basan}},
  \bibinfo{author}{\bibfnamefont{T.}~\bibnamefont{Risler}},
  \bibinfo{author}{\bibfnamefont{J.-F.} \bibnamefont{Joanny}},
  \bibinfo{author}{\bibfnamefont{X.}~\bibnamefont{Sastre-Garau}},
  \bibnamefont{and} \bibinfo{author}{\bibfnamefont{J.}~\bibnamefont{Prost}},
  \bibinfo{journal}{HFSP Journal} \textbf{\bibinfo{volume}{3}},
  \bibinfo{pages}{265} (\bibinfo{year}{2009}).

\bibitem[{\citenamefont{Montel et~al.}(2011)\citenamefont{Montel, Delarue,
  Elgeti, Malaquin, Basan, Risler, Cabane, Vignjevic, Prost, Cappello
  et~al.}}]{Montel2011}
\bibinfo{author}{\bibfnamefont{F.}~\bibnamefont{Montel}},
  \bibinfo{author}{\bibfnamefont{M.}~\bibnamefont{Delarue}},
  \bibinfo{author}{\bibfnamefont{J.}~\bibnamefont{Elgeti}},
  \bibinfo{author}{\bibfnamefont{L.}~\bibnamefont{Malaquin}},
  \bibinfo{author}{\bibfnamefont{M.}~\bibnamefont{Basan}},
  \bibinfo{author}{\bibfnamefont{T.}~\bibnamefont{Risler}},
  \bibinfo{author}{\bibfnamefont{B.}~\bibnamefont{Cabane}},
  \bibinfo{author}{\bibfnamefont{D.}~\bibnamefont{Vignjevic}},
  \bibinfo{author}{\bibfnamefont{J.}~\bibnamefont{Prost}},
  \bibinfo{author}{\bibfnamefont{G.}~\bibnamefont{Cappello}},
  \bibnamefont{et~al.}, \bibinfo{journal}{Phys. Rev. Lett.}
  \textbf{\bibinfo{volume}{107}}, \bibinfo{pages}{188102}
  (\bibinfo{year}{2011}).

\bibitem[{\citenamefont{Yang et~al.}(2014)\citenamefont{Yang, Manning, and
  Marchetti}}]{XY2014}
\bibinfo{author}{\bibfnamefont{X.}~\bibnamefont{Yang}},
  \bibinfo{author}{\bibfnamefont{M.~L.} \bibnamefont{Manning}},
  \bibnamefont{and} \bibinfo{author}{\bibfnamefont{M.~C.}
  \bibnamefont{Marchetti}}, \bibinfo{journal}{Soft Matter} p.
  \bibinfo{pages}{DOI: 10.1039/c4sm00927d} (\bibinfo{year}{2014}).

\bibitem[{\citenamefont{Takatori et~al.}(2014)\citenamefont{Takatori, Yan, and
  Brady}}]{Takatori2014}
\bibinfo{author}{\bibfnamefont{S.~C.} \bibnamefont{Takatori}},
  \bibinfo{author}{\bibfnamefont{W.}~\bibnamefont{Yan}}, \bibnamefont{and}
  \bibinfo{author}{\bibfnamefont{J.~F.} \bibnamefont{Brady}},
  \bibinfo{journal}{Phys. Rev. Lett.} \textbf{\bibinfo{volume}{113}},
  \bibinfo{pages}{028103} (\bibinfo{year}{2014}).

\bibitem[{\citenamefont{Solon et~al.}(2015)\citenamefont{Solon, Fily, Baskaran,
  Cates, Kafri, Kardar, and Tailleur}}]{Solon2015}
\bibinfo{author}{\bibfnamefont{A.~P.} \bibnamefont{Solon}},
  \bibinfo{author}{\bibfnamefont{Y.}~\bibnamefont{Fily}},
  \bibinfo{author}{\bibfnamefont{A.}~\bibnamefont{Baskaran}},
  \bibinfo{author}{\bibfnamefont{M.~E.} \bibnamefont{Cates}},
  \bibinfo{author}{\bibfnamefont{Y.}~\bibnamefont{Kafri}},
  \bibinfo{author}{\bibfnamefont{M.}~\bibnamefont{Kardar}}, \bibnamefont{and}
  \bibinfo{author}{\bibfnamefont{J.}~\bibnamefont{Tailleur}},
  \bibinfo{journal}{Nat. Phys.} pp. \bibinfo{pages}{11, 673--678}
  (\bibinfo{year}{2015}).

\bibitem[{\citenamefont{Barton et~al.}(2016)\citenamefont{Barton, Henkes,
  Weijer, and Sknepnek}}]{Rastko2016}
\bibinfo{author}{\bibfnamefont{D.~L.} \bibnamefont{Barton}},
  \bibinfo{author}{\bibfnamefont{S.}~\bibnamefont{Henkes}},
  \bibinfo{author}{\bibfnamefont{C.~J.} \bibnamefont{Weijer}},
  \bibnamefont{and} \bibinfo{author}{\bibfnamefont{R.}~\bibnamefont{Sknepnek}},
  \bibinfo{journal}{bioRxiv}  (\bibinfo{year}{2016}).

\bibitem[{\citenamefont{Henkes et~al.}(2011)\citenamefont{Henkes, Fily, and
  Marchetti}}]{Silke2011}
\bibinfo{author}{\bibfnamefont{S.}~\bibnamefont{Henkes}},
  \bibinfo{author}{\bibfnamefont{Y.}~\bibnamefont{Fily}}, \bibnamefont{and}
  \bibinfo{author}{\bibfnamefont{M.~C.} \bibnamefont{Marchetti}},
  \bibinfo{journal}{Phys. Rev. E} pp. \bibinfo{pages}{84, 040301}
  (\bibinfo{year}{2011}).

\bibitem[{\citenamefont{Fily and Marchetti}(2012)}]{Fily2012}
\bibinfo{author}{\bibfnamefont{Y.}~\bibnamefont{Fily}} \bibnamefont{and}
  \bibinfo{author}{\bibfnamefont{M.~C.} \bibnamefont{Marchetti}},
  \bibinfo{journal}{Phys. Rev. Lett.} \textbf{\bibinfo{volume}{108}},
  \bibinfo{pages}{235702 [5pages]} (\bibinfo{year}{2012}).

\bibitem[{\citenamefont{Su and Lan}(2016)}]{Su2016}
\bibinfo{author}{\bibfnamefont{T.}~\bibnamefont{Su}} \bibnamefont{and}
  \bibinfo{author}{\bibfnamefont{G.}~\bibnamefont{Lan}},
  \bibinfo{journal}{arXiv:1610.04254}  (\bibinfo{year}{2016}).

\bibitem[{\citenamefont{NESTOR-BERGMANN
  et~al.}(2016)\citenamefont{NESTOR-BERGMANN, GODDARD, WOOLNER, and
  JENSEN}}]{Jensen2016}
\bibinfo{author}{\bibfnamefont{A.}~\bibnamefont{NESTOR-BERGMANN}},
  \bibinfo{author}{\bibfnamefont{G.}~\bibnamefont{GODDARD}},
  \bibinfo{author}{\bibfnamefont{S.}~\bibnamefont{WOOLNER}}, \bibnamefont{and}
  \bibinfo{author}{\bibfnamefont{O.~E.} \bibnamefont{JENSEN}},
  \bibinfo{journal}{arXiv:1611.04744}  (\bibinfo{year}{2016}).

\bibitem[{\citenamefont{Yang et~al.}(2016)\citenamefont{Yang, Bi, Manning, and
  Marchetti}}]{XY2016}
\bibinfo{author}{\bibfnamefont{X.}~\bibnamefont{Yang}},
  \bibinfo{author}{\bibfnamefont{D.}~\bibnamefont{Bi}},
  \bibinfo{author}{\bibfnamefont{M.~L.} \bibnamefont{Manning}},
  \bibnamefont{and} \bibinfo{author}{\bibfnamefont{M.~C.}
  \bibnamefont{Marchetti}}, \bibinfo{journal}{(in preparation)}
  (\bibinfo{year}{2016}).

\bibitem[{\citenamefont{Bambardekar et~al.}(2015)\citenamefont{Bambardekar,
  Clement, Blanc, Chardes, and Lenne}}]{Kapil2014}
\bibinfo{author}{\bibfnamefont{K.}~\bibnamefont{Bambardekar}},
  \bibinfo{author}{\bibfnamefont{R.}~\bibnamefont{Clement}},
  \bibinfo{author}{\bibfnamefont{O.}~\bibnamefont{Blanc}},
  \bibinfo{author}{\bibfnamefont{C.}~\bibnamefont{Chardes}}, \bibnamefont{and}
  \bibinfo{author}{\bibfnamefont{P.-F.} \bibnamefont{Lenne}},
  \bibinfo{journal}{PNAS} \textbf{\bibinfo{volume}{112(5)}}
  (\bibinfo{year}{2015}).

\bibitem[{\citenamefont{Rauzzi et~al.}(2010)\citenamefont{Rauzzi, Lenne, and
  Lecuit}}]{Rauzi2010}
\bibinfo{author}{\bibfnamefont{M.}~\bibnamefont{Rauzzi}},
  \bibinfo{author}{\bibfnamefont{P.-F.} \bibnamefont{Lenne}}, \bibnamefont{and}
  \bibinfo{author}{\bibfnamefont{T.}~\bibnamefont{Lecuit}},
  \bibinfo{journal}{Nature} \textbf{\bibinfo{volume}{23}}
  (\bibinfo{year}{2010}).

\bibitem[{\citenamefont{Kasza et~al.}(2014)\citenamefont{Kasza, Farrell, and
  Zallen}}]{Zallen2014}
\bibinfo{author}{\bibfnamefont{K.~E.} \bibnamefont{Kasza}},
  \bibinfo{author}{\bibfnamefont{D.~L.} \bibnamefont{Farrell}},
  \bibnamefont{and} \bibinfo{author}{\bibfnamefont{J.~A.}
  \bibnamefont{Zallen}}, \bibinfo{journal}{PNAS}
  \textbf{\bibinfo{volume}{111(32):11732-7}} (\bibinfo{year}{2014}).

\bibitem[{\citenamefont{Hansen and McDonald}(2005)}]{Hansen2005}
\bibinfo{author}{\bibfnamefont{J.-P.} \bibnamefont{Hansen}} \bibnamefont{and}
  \bibinfo{author}{\bibfnamefont{I.}~\bibnamefont{McDonald}},
  \emph{\bibinfo{title}{Theory of simple liquids}}
  (\bibinfo{publisher}{Academic Press}, \bibinfo{year}{2005}),
  \bibinfo{edition}{3rd} ed.

\bibitem[{\citenamefont{Levashov et~al.}(2011)\citenamefont{Levashov, Morris,
  and Egami}}]{Egami2011}
\bibinfo{author}{\bibfnamefont{V.~A.} \bibnamefont{Levashov}},
  \bibinfo{author}{\bibfnamefont{J.~R.} \bibnamefont{Morris}},
  \bibnamefont{and} \bibinfo{author}{\bibfnamefont{T.}~\bibnamefont{Egami}},
  \bibinfo{journal}{Phys. Rev. Lett.} \textbf{\bibinfo{volume}{106}}
  (\bibinfo{year}{2011}).

\bibitem[{\citenamefont{Pan et~al.}(2016)\citenamefont{Pan, Heemskerk, Ibar,
  Shraiman, and Irvine}}]{Pan2016}
\bibinfo{author}{\bibfnamefont{Y.}~\bibnamefont{Pan}},
  \bibinfo{author}{\bibfnamefont{I.}~\bibnamefont{Heemskerk}},
  \bibinfo{author}{\bibfnamefont{C.}~\bibnamefont{Ibar}},
  \bibinfo{author}{\bibfnamefont{B.~I.} \bibnamefont{Shraiman}},
  \bibnamefont{and} \bibinfo{author}{\bibfnamefont{K.~D.}
  \bibnamefont{Irvine}}, \bibinfo{journal}{PNAS}
  \textbf{\bibinfo{volume}{113}}, \bibinfo{pages}{E6974}
  (\bibinfo{year}{2016}).

\bibitem[{\citenamefont{J et~al.}(2016)\citenamefont{J, BA, WJ, and
  H}}]{Zimmermann2016}
\bibinfo{author}{\bibfnamefont{Z.}~\bibnamefont{J}},
  \bibinfo{author}{\bibfnamefont{C.}~\bibnamefont{BA}},
  \bibinfo{author}{\bibfnamefont{R.}~\bibnamefont{WJ}}, \bibnamefont{and}
  \bibinfo{author}{\bibfnamefont{L.}~\bibnamefont{H}}, \bibinfo{journal}{PNAS}
  \textbf{\bibinfo{volume}{113}}, \bibinfo{pages}{2660} (\bibinfo{year}{2016}).

\end{thebibliography}



\subsection*{Supporting Information (SI)}
\subsubsection*{Traction-based Mechanical Inference}
\paragraph*{Force balance in the vertex model}
\begin{figure}[!h]
\begin{centering}
\includegraphics[width=1.00\columnwidth]{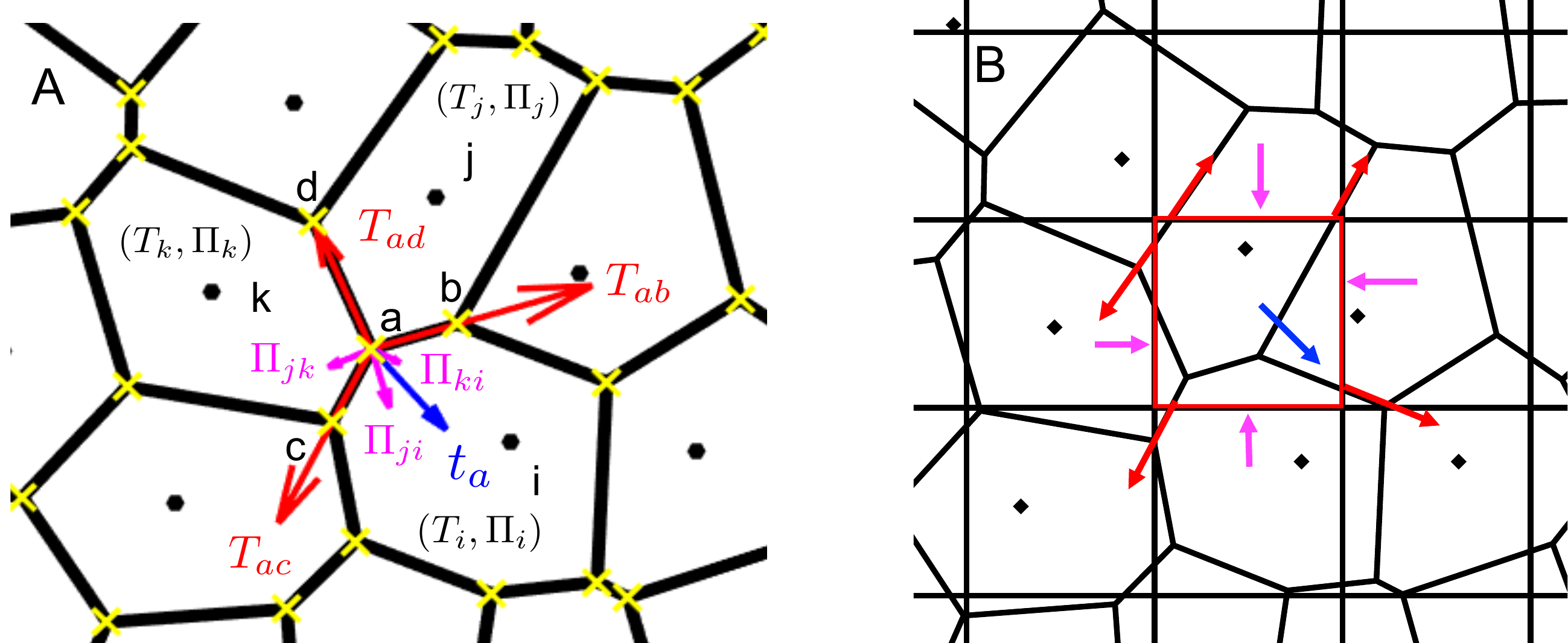}
\caption{A. Schematic of force balance at vertex $a$ joining the neighboring cells $i,j,k$, with tangential forces from edge tensions (red) and normal forces from pressure differences (magenta). The edge tensions at vertex $a$ are determined by the sum of the cortical tensions ($T_i,T_j,T_k$) of the neighboring cells (see Eq.~\eqref{TP1}). The normal forces are determined by the difference of the pressures ($\Pi_i, \Pi_j, \Pi_k$) of the neighboring cells (see Eq.~\eqref{TP2}). $\mathbf{t}_a$ is the traction force vertex $a$ exerts on the substrate. B. Schematic of force balance at the coarse-grained grid. {\color{black}The interaction force on the grid element (red arrow for tension and magenta arrow for pressure-induced normal force) balances the traction force (blue arrow) the grid element exerts on the substrate.}}
\label{fig:inference}
\end{centering}
\end{figure}
Force balance at each vertex requires
\begin{equation}
\bm t_a=\bm F_a, 
\label{eqn:force_balance_supp}
\end{equation}
where $\mathbf{t_a}$ is the traction force vertex $a$ exerts on the substrate, which is balanced by the interaction force $\bm F_a=-\nabla_a E$. We consider for concreteness four adjacent vertices $a,b,c,d$ and corresponding cells $i,j,k$, as shown in Fig.\ref{fig:inference}A.  
Using the additivity of the shape energy, we write
\begin{equation}
\bm F_a=-\bm\nabla_a E=-\left(\frac{\partial E_i}{\partial \bm r_a}+\frac{\partial E_j}{\partial \bm r_a}+\frac{\partial E_k}{\partial \bm r_a}\right)\;.
\label{force_balance_a}
\end{equation}
The right hand side of Eq.~\eqref{force_balance_a} can be decomposed  in terms of tensions and pressure jumps along and perpendicular to the cell edges, as illustrated in Fig.\ref{fig:inference}A, by writing
\begin{equation}
\frac{\partial E_i}{\partial\bm r_a}=-\Pi_i\frac{\partial A_i}{\partial \bm r_a}+T_i\frac{\partial P_i}{\partial\bm r_a}\;,
\label{dEdr}
\end{equation}
where $T_i$ and $\Pi_i$ are the cortical tension and hydrostatic pressure for cell $i$, respectively, defined as~\cite{Kevin2012}
\begin{equation}
T_i\equiv\frac{\partial E_i}{\partial P_i},~~~~~~\Pi_i\equiv-\frac{\partial E_i}{\partial A_i}.
\label{TP}
\end{equation}
The geometric factors in Eq.~\ref{dEdr} can be written as 
\begin{equation}
\frac{\partial A_i}{\partial \bm r_a}=(\hat{\bm n}_{ik}l_{ac}+\hat{\bm n}_{ij}l_{ab})/2,~~~~~\frac{\partial P_i}{\partial \bm r_a}=-(\hat{\bm l}_{ab}+\hat{\bm l}_{ac}),
\label{dAdr}
\end{equation}
where $\hat{\bm n}_{ik}$ is the unit vector perpendicular to edge $ac$ pointing from cell $i$ to cell $k$, $l_{ac}$ is the length of edge $ac$, and $\hat{\bm l}_{ad}$ is the unit vector along edge $ad$ pointing from vertex $a$ to vertex $d$. Combining Eq.~\eqref{eqn:force_balance_supp}-Eq.~\eqref{dAdr}, 
we obtain 
%
\begin{subequations}
\begin{gather}
\mathbf{t_a}=T_{ab}\hat{\bm l}_{ab}+T_{ac}\hat{\bm l}_{ac}+T_{ad}\hat{\bm l}_{ad}\\\notag +\frac{1}{2}\Pi_{ji}\hat{\bm n}_{ji}l_{ab}+\frac{1}{2}\Pi_{jk}\hat{\bm n}_{jk}l_{ad}+\frac{1}{2}\Pi_{ki}\hat{\bm n}_{ki}l_{ac}\;,
\label{decomposed}
\end{gather}
\end{subequations}
%
where 
\begin{eqnarray}
\label{TP1}
T_{ab}=T_i+T_j~~~T_{ac}=T_i+T_k~~~T_{ad}=T_j+T_k\\
\Pi_{ji}=\Pi_j-\Pi_i~~~\Pi_{jk}=\Pi_j-\Pi_k~~~\Pi_{ki}=\Pi_k-\Pi_i
\label{TP2}
\end{eqnarray}
are edge tensions and pressure jumps along and across cell edges, respectively. Eq.~\eqref{decomposed}-Eq.~\eqref{TP2} form the basis for the traction-based mechanical inference and can be written in a general form as
\begin{equation}
\mathbf{F}_a(\{\Pi_i\},\{T_{i}\})=\mathbf{t}_a\;.
\label{force_balance_SI}
\end{equation}
Operationally, we write the force balance equation in a matrix form
\begin{equation}
\bm M\cdot \bm \phi=\bm t\;,
\label{inference_mat}
\end{equation}
where $\bm M$ is a $4N\times 2N$ structural matrix incorporating the orientation and length of the edge vectors, $\bm \phi$ is a $2N\times 1$ column vector whose elements are the cortical tensions and pressures, and $\bm t$ is the $4N\times 1$ column vector of the traction forces at each vertex. The tensions and pressures are obtained by inverting the structural matrix
\begin{equation}
\bm \phi=\bm M^{-1}\cdot \bm t,
\label{inverse}
\end{equation}
which is equivalent to a least square minimization of 
\begin{equation}
Tr[(\bm M\cdot\bm \phi-\bm t)^T\cdot (\bm M\cdot\bm \phi-\bm t)]\;.
\label{minimization}
\end{equation}
In practice, this is done with the Moore-Penrose pseudoinverse algorithm~\cite{Kevin2012}. 
{\color{black}\paragraph*{Force balance on the coarse-grained grid}
In real tissues, the traction force acts not only at cell vertices, but also at focal adhesions and throughout the basal area of a cell, and can be measured using TFM on a coarse-grained grid. Therefore, we generalize the traction-based mechanical inference by coarse-graining the tissue on a Cartesian grid and impose force balance on each grid element (Fig.\ref{fig:inference}B)
\begin{equation}
\mathbf{F}_{grid}(\{\Pi_i\},\{T_{i}\})=\mathbf{t}_{grid}\;,
\label{force_balance_SI}
\end{equation}
{\color{black}where $\mathbf{F}_{grid}$ is the interaction force on each grid element as calculated from tensions and pressures as shown in Fig.\ref{fig:inference}B, and $\mathbf{t}_{grid}$ is the traction force the grid element exerts on the substrate. A grid element is a closed region of the tissue. In a Cartesian grid, it is a square element as highlighted in Fig.\ref{fig:inference}B. The interaction force acts on the grid element through the boundary and is generally expressed as a surface integral, or a line integral in 2D, as
\begin{equation}
\mathbf{F}_{grid}=\int_{\partial S}\left[T_{ab}\hat{\bm n}_{ab}\delta(l-l_s)+\Pi_i\hat{\bm n}_s\right]dl\;,
\label{eqn:grid_interaction}
\end{equation}
where tension is a point force concentrated at the intersection $l_s$ between the cell edge and the grid element boundary, while pressure induces a normal force in the normal direction  $\hat{\bm n}_s$ of the boundary. Notice that the pressure varies as the boundary crosses different cells. Using cell segmentation data together with Eq.~\eqref{eqn:grid_interaction} and Eq.~\eqref{TP1}, the interaction force can be written explicitly in terms of cortical tensions and cell pressures. The traction force the grid element exerts on the substrate can be measured using TFM. By inverting the force balance Eq.~\eqref{force_balance_SI} using the Moore-Penrose pseudoinverse algorithm, one can obtain tensions and pressures, from which the local stress tensor can be constructed using Eq.~\eqref{eqn:cellular_stress} in the main text.

To test the coarse-grained mechanical inference, we generate cell network configurations using the SPV model. The traction force of the grid element is calculated as the sum of tensions (normal forces) on the vertices (cell edges) within the grid element. It is numerically verified that the traction force is exactly balanced by the interaction force for each grid element. We have tested the coarse-grained mechanical inference in our numerical simulation with a grid element size of the order of a single cell size. The inferred stress agrees perfectly with the accurate value obtained from the cell shape using the energy functional of the tissue (Fig.\ref{fig:distribution} in the main text).} 

This coarse-grained mechanical inference requires knowledge of cell shapes as well as traction forces. The cell shape allows one to express the interaction force in terms of tension and pressure and resolves the stress down to a single cell level. The resolution of the coarse-graining can be adjusted to adapt to the resolution of the traction force microscopy, where a higher TFM resolution will yield a more accurate result for the inferred tensions and pressure differences by over-constraining the system. A systematic study of the method will be published elsewhere~\cite{XY2016}.

The traction-based mechanical inference does not require knowledge of the energy functional or the material property of the tissue, and is based on the assumption of force balance and the decomposition of interaction forces into junctional tensions and cell pressures. We expect our method to have general applicability to epithelial tissues where cell shape and traction forces can be measured simultaneously.}
\subsubsection*{Tissue Stress}
\begin{figure}[!h]
\begin{centering}
\includegraphics[width=1.00\columnwidth]{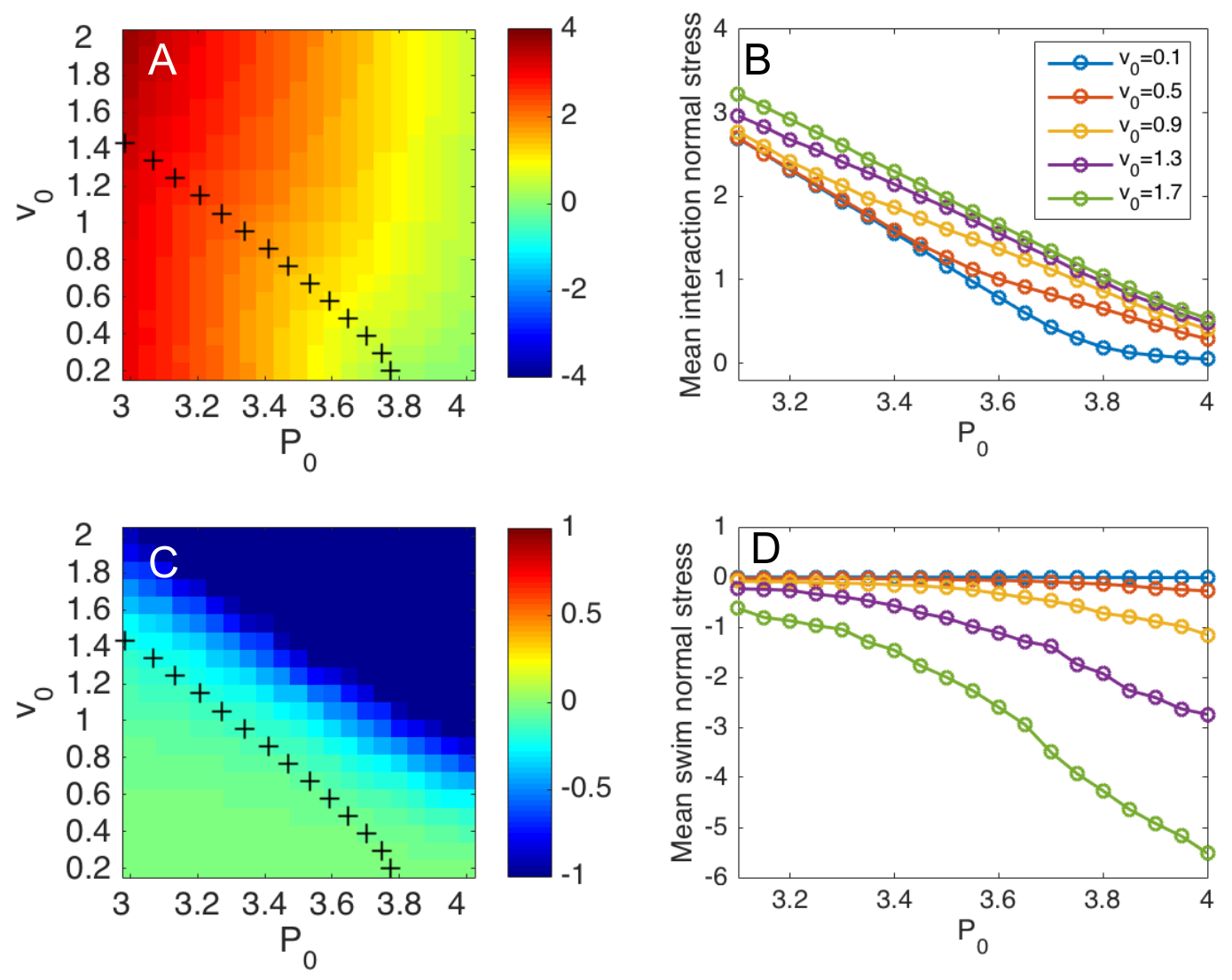}
\caption{ A(C): heat map of the mean interaction (swim) normal stress of the tissue in the $(v_0,P_0)$ plane. The black crosses outline the solid-liquid phase boundary determined by $q=3.813$. Red indicates contractile stress and blue extensile stress. B(D): mean interaction (swim) normal stress  as a  function of $P_0$ for various $v_0$ ($400$ cells for $T=1000$ and $D_r=0.1$).}
\label{fig:int_swim}
\end{centering}
\end{figure}

From cellular tensions and pressures one can also evaluate the mean interaction stress for the whole tissue, given by 
\begin{equation}
\sigma_{\alpha\beta}^{int}=-\Pi_T\delta_{\alpha\beta}+\frac{1}{A_T}\sum_{ab}T_{ab}^{\alpha}l_{ab}^{\beta},
\label{tissue_stress}
\end{equation}
where the summation is over all edges, $A_T$ is the area of the tissue, and the pressure conjugated to the tissue area $A_T$ is given by
\begin{equation}
\Pi_T  = \sum_i \Pi_i \frac{A_i}{A_T}.
\label{additive}
\end{equation}
Combining Eq.~\eqref{eqn:cellular_stress} in the main text with Eq.~\eqref{tissue_stress}-Eq.~\eqref{additive}, we show that the interaction tissue stress is related to the cellular stress as
\begin{equation}
\sigma_{\alpha\beta}^{int} = \frac{1}{A_T}\sum_{i} A_i\sigma^{(i)int}_{\alpha\beta}.
\label{additive_stress}
\end{equation}
Next we show that for the energy functional given by Eq.~\eqref{eqn:energy_square} of the main text, the tissue interaction normal stress can be expressed in terms of cell areas and perimeters. We first focus on the interaction normal stress of cell $i$ defined as 
\begin{equation}
\sigma_n^{(i)int} = \frac{1}{2}\left(\sigma_{xx}^{(i)int}+\sigma_{yy}^{(i)int}\right).
\label{eqn:cellular_normal}
\end{equation}
Using Eq.~\eqref{eqn:cellular_stress} in the main text, we rewrite Eq.~\eqref{eqn:cellular_normal} as
\begin{equation}
\sigma_n^{(i)int} = -\Pi_i + \frac{1}{4A_i}\sum_{j\in n.o.i}\left[(T_i+T_j)l_{ij}\right],
\label{eqn:cellular_normal_2}
\end{equation}
where we used the relation in Eq.~\eqref{TP1} and the summation is over the neighbors of cell $i$. $l_{ij}$ is the length of the edge shared by cell $i$ and cell $j$. Using the energy functional (Eq.~\eqref{eqn:energy_square} in the main text) and the definitions for tension and pressure (Eq.~\eqref{TP}), we have
\begin{equation}
\sigma_n^{(i)int} = 2K_A(A_i-A_0)+\frac{K_P}{2A_i}\sum_{j\in n.o.i}\left[(P_i-P_0)l_{ij}+(P_j-P_0)l_{ij}\right].
\label{eqn:cellular_normal_3}
\end{equation}
Substituting Eq.~\eqref{eqn:cellular_normal_3} into Eq.~\eqref{additive_stress}, we obtain the interaction normal stress for the tissue in terms of cell areas and perimeters
\begin{equation}
\sigma_n^{int} = \frac{1}{A_T}\sum_i\left[2K_A A_i(A_i-A_0) + K_P P_i(P_i-P_0)\right],
\label{eqn:tissue_normal}
\end{equation}
where we used the identity
\begin{subequations}
\begin{gather}
\sum_{j\in n.o.i}l_{ij} = P_i\\
\sum_i\sum_{j\in n.o.i}(P_j-P_0)l_{ij} = \sum_{i}(P_i-P_0)P_i.
\end{gather}
\end{subequations}
Eq.~\eqref{eqn:tissue_normal} expresses the interaction tissue stress in terms of cell shape parameters, providing a prescription for extracting mechanical information directly from cell images. We plot the mean normal interaction stress (Eq.~\eqref{eqn:tissue_normal}) and the mean normal swim stress (Eq.~\eqref{tissue_level_stress}) as a function of $v_0$ and $P_0$ in Fig.~\ref{fig:int_swim}. Notice that the interaction stress is contractile and the swim stress is extensile. The latter vanishes in the solid state and dominates deep in the liquid state, resulting in a change of sign of the total normal stress as shown in Fig.~\ref{fig:normal_stress} in the main text. 
\subsubsection*{Effect of $A_0$ on cell dynamics}
\begin{figure}[!h]
\begin{centering}
\includegraphics[width=0.80\columnwidth]{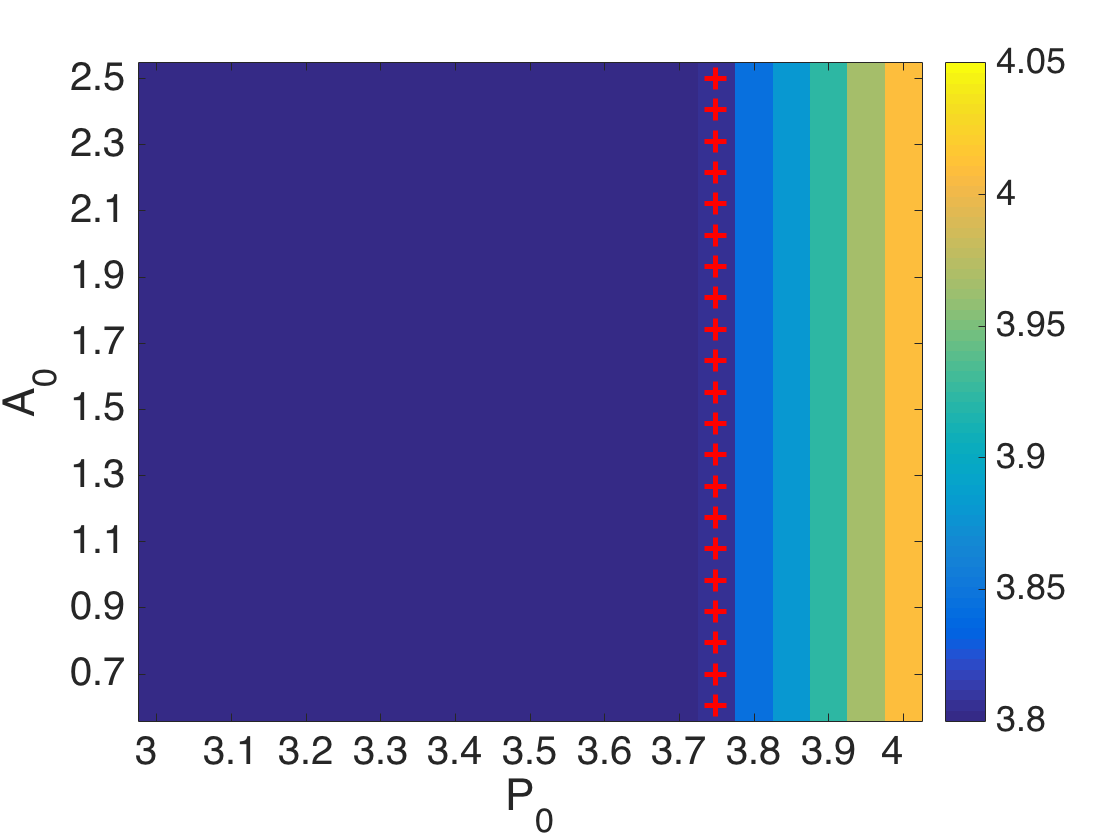}
\caption{Phase diagram in the $A_0-P_0$ plane at $v_0=0.2$. The red crosses outline the solid-liquid phase boundary determined by $q=3.813$. Notice that $A_0$ does not affect the liquid-solid phase transition. The simulation is performed using the SPV model for a duration of $T=1000$ with $400$ cells at $D_r=1.0$ with periodic boundary condition.}
\label{fig:A0}
\end{centering}
\end{figure}
For a system with periodic boundary conditions and fixed size, the parameter $A_0$ plays no essential role in the cell dynamics. It only renormalizes the pressure of the system by a constant, but does not affect cell shapes, forces on vertices, or the anisotropic part of the stress tensor.  To see this, we rewrite the energy of the system using the average cell area $\bar A$.  Up to a constant offset, the total energy reads:
\begin{subequations}
\begin{gather}
	E = \sum_{i=1}^N{\Big[K_A(A_i-\bar A)^2 + K_P(P_i-P_0)^2\Big]} \\\notag
		+ \frac{1}{N}K_AA_T^2 - 2K_AA_TA_0\text{.}
\label{eqn:total_energy}
\end{gather}
\end{subequations}
Note that the parameter $A_0$ only appears in the last term of Eq.~\eqref{eqn:total_energy}, which merely offsets the pressure of the system by $2K_AA_0$.  Thus, all forces $\bm{F}_i$ and shear stresses are independent of $A_0$. The liquid-solid phase transition is unaffected by $A_0$ as shown in Fig.\ref{fig:A0}.

{\color{black}\subsubsection*{Effect of Voronoi constraint on the tissue stress}
\begin{figure}[!h]
\begin{centering}
\includegraphics[width=1.00\columnwidth]{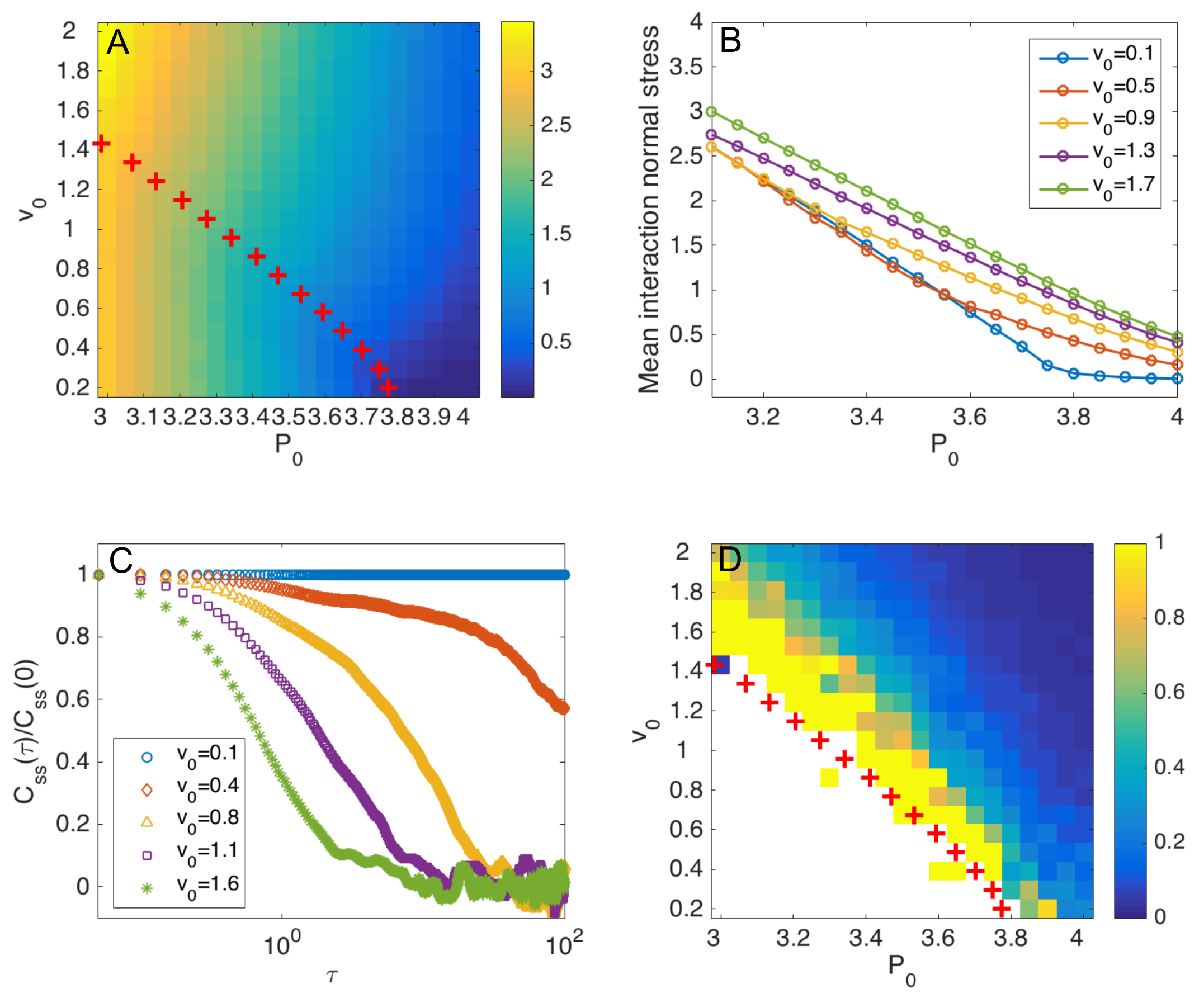}
\caption{Behavior of normal and shear tissue stress with Voronoi constraint from the energy derivative with respect to imposed strain (Eq.~\eqref{eqn:SPV_stress_ENG}). A. Heat map of the mean interaction normal stress as a function of $v_0$ and $P_0$. The red crosses outline the solid-liquid phase boundary determined by $q=3.813$. B. mean interaction normal stress as a function of $P_0$ at various $v_0$. C. Time autocorrelation of the mean interaction shear stress for various $v_0$ at $P_0=3.45$. D. heat map of the effective viscosity (Eq.~\eqref{eqn:green_kubo} in the main text). The simulation is performed with 100 cells and a duration of $T=1000$ with periodic boundary condition.}
\label{fig:SPV_ENG}
\end{centering}
\end{figure}
The SPV model can be mapped to an active vertex model (AVM) with effective constraints imposed by the Voronoi tessellation on the vertex positions. In a vertex model free of Voronoi constraint, the interaction stress tensor $\sigma_{\alpha\beta}^{int}$ as defined by Eq.~\eqref{tissue_level_stress} in the main text represents precisely the stress acting on the periodic boundary. This is not, however, the case for the SPV model, because the creation of cell shapes using a Voronoi tessellation adds constraints to a vertex model. In this section, we show that the results presented in the paper are robust against the Voronoi constraint. 

The normal stress as defined in Eq.~\eqref{eqn:stress_component} in the main text is unaffected by the Voronoi constraint, but the shear stress does not correspond to the shear stress at the boundary of the system. This is because an affine shear deformation of all cell positions does not induce an affine shear transformation of all vertex positions. Indeed, the $2N$ vertices would in total provide $4N$ degrees of freedom, but the Voronoi tessellation reduces this to the $2N$ degrees of freedom of the $N$ cell positions.  

In the SPV model, the stress experienced by the boundary can be computed by quantifying the total energy change induced by a deformation of the periodic box described by the strain tensor $G_{\alpha\beta}$ while affinely displacing all cell positions and respecting the Voronoi tessellation:
\begin{equation}
\tilde\sigma_{\alpha\beta}^{int} = \frac{1}{A_T}\frac{d E}{d G_{\alpha\beta}},
\label{eqn:SPV_stress_ENG}
\end{equation}
where the tilde is used to denote the stress at the boundary under Voronoi constraint. The shear component of this tensor is different from that of the stress $\sigma_{\alpha\beta}^{int}$ defined in Eq.~\eqref{tissue_level_stress} in the main text.

To numerically study how different $\tilde\sigma_{\alpha\beta}^{int}$ and $\sigma_{\alpha\beta}^{int}$ typically are, we impose a sufficiently small strain on the cell positions and calculate the tissue energy difference before and after the imposed strain. To incorporate the Voronoi constraint, the cell network is constructed based on the cell positions before and after the imposed strain according to the Voronoi tessellation. An expansion along $x$ and $y$ axis is implemented to compute the normal components of the stress $\tilde{\sigma}_{xx}^{int}$ and $\tilde{\sigma}_{yy}^{int}$, using
\begin{equation}
\tilde{r}_x^{i} = r_x^i(1+\epsilon),~~~~\tilde{r}_y^{i} = r_y^i(1+\epsilon),
\label{eqn:expansion}
\end{equation}
respectively. Here, $\bm r$ is the cell position and $\epsilon\ll1$ is the imposed strain. Similarly, simple shear deformations along $x$ and $y$ axes are implemented to compute the off-diagonal components of the stresses $\tilde{\sigma}_{xy}^{int}$ and $\tilde{\sigma}_{yx}^{int}$, using the transformations
\begin{equation}
\tilde{r}_x^{i} = r_x^i+\epsilon r_y^i,~~~~\tilde{r}_y^{i} = r_y^i+\epsilon r_x^i,
\label{eqn:imposed_shear}
\end{equation}
respectively.
 
We present the results of the stress at the boundary under Voronoi constraint in Fig.\ref{fig:SPV_ENG}. Comparing with the results in the main text, the effect of the Voronoi tessellation does not affect the behavior of the normal and the shear stress as presented in the main text. This suggests that for practical purposes, one can apply the definitions of stress of the vertex model to approximate the stress in the SPV model. The insensitivity of the behavior of the stress to the geometrical constraint also implies general applicability of the SPV model in describing the mechanics of confluent epithelial tissue in experiment.}

{\color{black}\subsubsection*{Autocorrelation of traction forces}
\begin{figure}[!h]
\begin{centering}
\includegraphics[width=1.00\columnwidth]{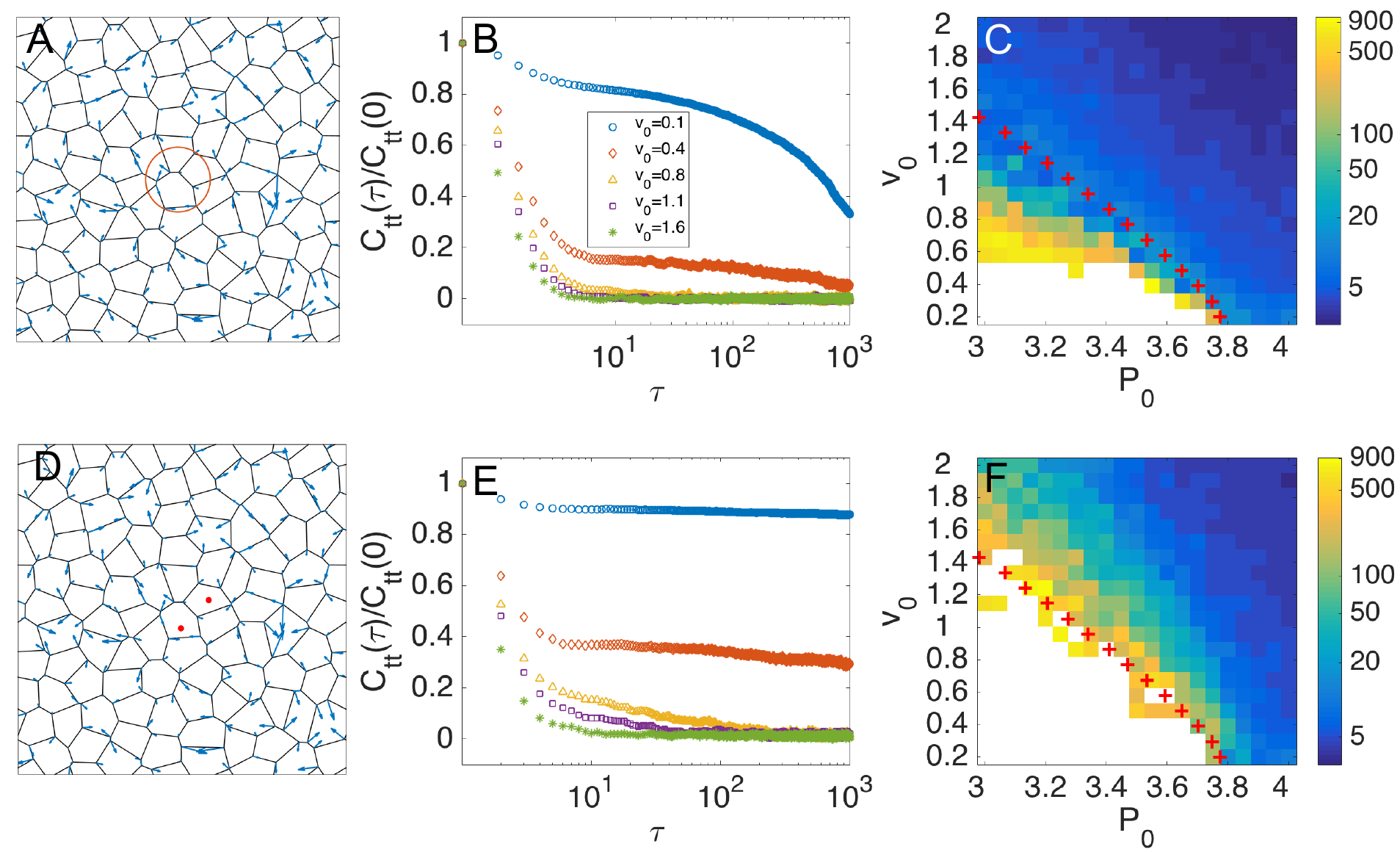}
\caption{The time autocorrelation of traction force slows down as the tissue transits from fluid state to solid state. A-C (D-F) correspond to the autocorrelation evaluated in the Eulerian (Lagrangian) reference frame. A: vertex tractions are averaged within a circle of radius $R=0.1L$ fixed in space, where $L$ is the length of the simulation box. {\color{black}D: average tractions at vertices of cells initially located within a circle of radius $R=0.1L$ and then tracked in time (the corresponding cells are labeled in red).} B and E: Normalized autocorrelation of traction force at  $P_0=3.45$ in the Eulerian (B) and Lagrangian (E) frame. C and F: Heat maps of the correlation time $\tau_m$ in the Eulerian (C) and Lagrangian (F) frame. The white region corresponds to correlation time beyond the duration of the simulation. The simulation is performed over a total time $T=10000$ for $100$ cells with $D_r=1.0$ and periodic boundary condition.}
\label{fig:traction_corr}
\end{centering}
\end{figure}
The slowing down of the autocorrelation of the shear stress implies a similar behavior for the autocorrelation of traction forces due to the force balance condition $t_{\alpha} = \partial_{\beta} \sigma_{\alpha\beta}^{int}$. We have shown this explicitly by evaluating the autocorrelation function of the vertex tractions averaged over a small patch of tissue as a function of $v_0$ and $P_0$, as shown in Fig.~\ref{fig:traction_corr}. In contrast to the shear stress, we cannot average the traction over the entire system, as due to force balance the net traction force is zero. Instead we compute the two-time correlation function of the traction forces averaged over a circular patch of tissue of radius $R$. Such a mean local traction is calculated in two reference frames: the Eulerian frame where the circular patch of radius $R=0.1L$ is fixed in space (Fig.~\ref{fig:traction_corr}A) and the Lagrangian reference frame where cells initially within the patch are tracked in time (Fig.~\ref{fig:traction_corr}D). To reduce the finite size fluctuations, we compute the autocorrelation using $7$ patches at different positions and average over the patches. Experimental traction forces are usually measured in an Eulerian reference frame by TFM. Such measurements could, however, be transformed to a Lagrangian reference frame if cell tracking data are available.

Figure~\ref{fig:traction_corr} B and E show the autocorrelation of the traction,
$C_{tt}(\tau) = \langle t_x(t_0)t_x(t_0+\tau)\rangle_{t_0}$, with $t_x$ the $x$ component of the the average traction in the Eulerian and Lagrangian frame, respectively. When calculated in the Lagrangian frame, the correlations show the same dramatic slow down when the solid is approached from the liquid side as found in the autocorrelation of the mean shear stress. In the Eulerian frame, however, a second time scale associated with cells moving in and out of the fixed patch masks the slowing down of the relaxation at long times. The same behavior is obtained when calculating correlations of shear stress averaged over a local patch instead of over the entire system. 
The slowing down of the relaxation is quantified in Fig.~\ref{fig:traction_corr} C and F that show heat maps of the correlation time $\tau_m$ defined as $C_{tt}(\tau_m)/C_{tt}(0)=0.02$ in the Eulerian and Lagrangian frames, respectively. 
The correlation time increases as the system transits from the fluid state to the solid state in both reference frames and diverges deep in the solid state. The behavior is less clear in the Eulerian reference frame, where cell transition across the patch introduces a new time scale that masks the divergence.  
In summary, this confirms that the divergence of traction and shear relaxation can in principle be seen in TFM data, especially if combined with
cell tracking so that correlations can be computed in a Lagrangian frame. Our results suggest a new way of analyzing TFM data not previously attempted to provide rheological information about the tissue.}

\end{document}